\pdfoutput=1
\newif\ifFull
\Fullfalse
\ifFull
\documentclass[11pt]{article}
\usepackage[margin=1in]{geometry}
\else
\documentclass{sig-alternate-05-2015}
\fi
\usepackage{cite,microtype,graphicx}
\usepackage{tikz}
\usetikzlibrary{positioning}
\usepackage{subfig}
\usepackage{floatrow}
\usepackage{amssymb}
\usepackage{amsmath}
\usepackage[ruled]{algorithm2e}
\usepackage{hyperref} 
\usepackage{array} 
\begin{document}

\CopyrightYear{2016}
\setcopyright{acmlicensed}
\conferenceinfo{SIGSPATIAL'16,}{October 31-November 03, 2016,
Burlingame, CA, USA}
\isbn{978-1-4503-4589-7/16/10}\acmPrice{\$15.00}
\doi{http://dx.doi.org/10.1145/2996913.2996976}

\title{A Topological Algorithm for Determining How \\
          Road Networks Evolve Over Time}

\numberofauthors{3}
\author{
\alignauthor
       Michael T. Goodrich\\
       \affaddr{Dept. of Computer Science}\\
       \affaddr{Univ. of California, Irvine}\\
       \affaddr{Irvine, CA 92697 USA}\\
       \email{goodrich@uci.edu}
\alignauthor
       Siddharth Gupta\\
       \affaddr{Dept. of Computer Science}\\
       \affaddr{Univ. of California, Irvine}\\
       \affaddr{Irvine, CA 92697 USA}\\
       \email{guptasid@uci.edu}
\alignauthor
       Manuel R. Torres\\
       \affaddr{Dept. of Computer Science}\\
       \affaddr{Univ. of California, Irvine}\\
       \affaddr{Irvine, CA 92697 USA}\\
       \email{mrtorres@uci.edu}
}

\date{}

\maketitle
\pagestyle{plain}
\def\thepage{\arabic{page}}

\begin{abstract}
We provide an efficient algorithm for determining how a road network has
evolved over time, given two snapshot instances from different dates.
To allow for such determinations across different databases and 
even against hand-drawn maps, we take a strictly topological
approach in this paper, so that we compare road networks
based strictly on graph-theoretic properties.
Given two road networks of same region from two different dates, 
our approach allows one to match 
road network portions that remain intact and also point out 
added or removed portions.
We analyze our algorithm both theoretically, showing that it runs in 
polynomial time for non-degenerate 
road networks even though a related problem is NP-complete,
and experimentally, using dated road networks from the TIGER/Line
archive of the U.S.~Census Bureau.
\end{abstract}

\ccsdesc[500]{Information systems~Geographic information systems}

\printccsdesc

\keywords{map evolution; isomorphism; conformal matching}


\section{Introduction}
\begin{figure}[hbt!]
\centering
\includegraphics[width=.8\columnwidth, trim= 1in 3.1in 1in 2.5in, clip]{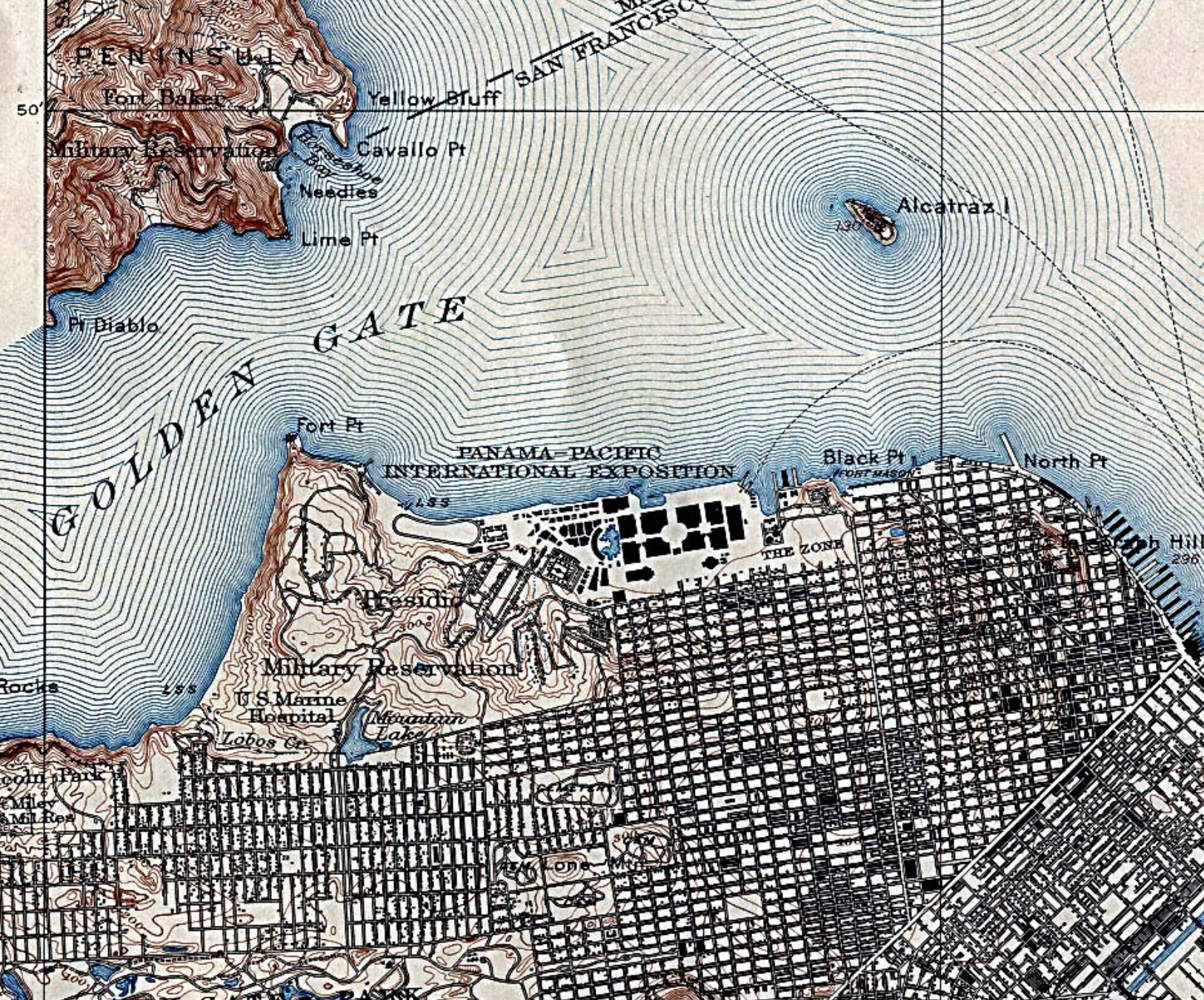} 
\\[0.1in]
\includegraphics[width=.8\columnwidth, trim= 1in 2.5in 1in 2.5in, clip]{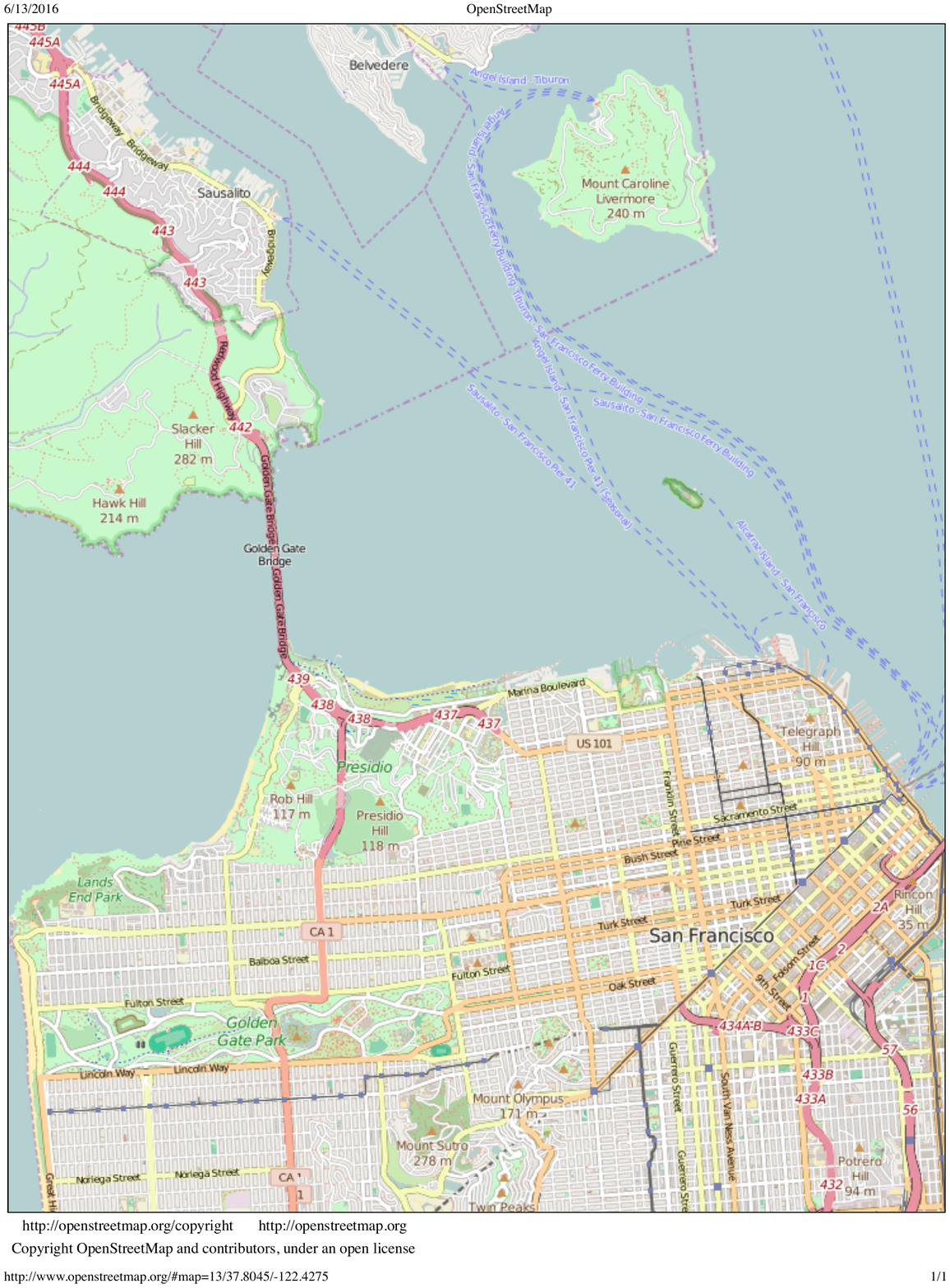} 
\caption{A map of San Francisco from 1915 and one from 2016 (taken 
from OpenStreetMap).
The top image is in the public domain; the bottom image is licensed under
the Open Database License, CC BY-SA.
Note that most of the roads in both maps are not labeled.
}
\label{fig:sf}
\end{figure}

Road network algorithms
are an important topic of study in Geographic Information
Systems (GIS), in that road networks facilitate transportation 
and are the products of social, geographic, economic, and political forces.
In addition, road networks are interesting data types, in that they combine
both geometric information and graph-theoretic information.
\ifFull
\else
(E.g., see~\cite{Eppstein:2008}.)
\fi
Formally, we view a road networks as a graph, 
where we create a vertex for every
road intersection or major jog, 
and we create an edge for every pair of such vertices that have a road
segment that joins them. 
In addition, some road networks are annotated with geometric/geographic
information, such as the GPS coordinates of some vertices or 
labels identifying road names.
Nevertheless, because road networks may contain 
many vertices and edges without such geometric/geographic
information, we are interested in this paper in studying road networks from
strictly a topological viewpoint, that is, as embedded graphs.
Specifically,
we are interested in the problem of determining how road
networks evolve over time,
e.g., highlighting places
where new roads and bridges are added and where old roads and bridges are
removed.
(See Figure~\ref{fig:sf}.)

\pagebreak
\subsection{Problem Definition}
Viewed topologically in terms of their graph properties,
road networks are embedded graphs,
that is,
the edges incident on each vertex are
given in a particular order (i.e., clockwise or counterclockwise),
which defines a topological structure
for the graph
known as a \emph{rotation system} (e.g., see~\cite{JGT:JGT3190020402}).
Road networks are not typically planar graphs (e.g., see~\cite{Eppstein:2008}),
however, since there are edge crossings, for example, at overpasses.
Thus, we cannot in general apply algorithms for planar graphs to road networks.
Nevertheless, the vertices in road networks have bounded degrees
(since the number of roads that meet at a single junction cannot 
be arbitrarily large); hence, a road network with $n$ vertices has
$O(n)$ edges.

Given two undirected graphs, $G_1$ and $G_2$, 
an \emph{isomorphism} of $G_1$ and $G_2$ is a bijection, $f$, from the vertices
of $G_1$ to the vertices of $G_2$ such that $(u,v)$ is an edge in
$G_1$ if and only
if $(f(u),f(w))$ is an edge in $G_2$ (e.g., see~\cite{McKay201494}).
In the \emph{subgraph isomorphism} problem, we are given two graphs,
$G_1$
and $G_2$, and asked to determine whether there is 
a subgraph of $G_1$ isomorphic to $G_2$.
This problem is NP-complete, even if $G_1$ is an embedded planar graph,
by a reduction from 
the planar Hamiltonian circuit problem~\cite{doi:10.1137/0205049}.
Thus, defining the best matching 
between two road networks simply in terms of a maximum common subgraph
is unlikely to lead to a polynomial-time algorithm.
So let us restrict the types of matchings we consider.

Suppose we are given a subgraph, $G_1'$, of a graph, $G_1$, and 
a subgraph, $G_2'$ of a graph, $G_2$, such that
$G_1$ and $G_2$ are embedded graphs, i.e., having specified rotation systems.
Suppose further that $f$ is an isomorphism from $G_1'$ to $G_2'$.
We say that $f$ is \emph{conformal} if it satisfies
the following conditions:
\begin{enumerate}
\item
For every vertex $v$ in $G_1'$,
$v$ has the same degree in $G_1$ as $f(v)$ has in $G_2$.
That is, we only match vertices having the same degree.
\item
For every pair of incident edges, $(v,u)$ and $(v,w)$, in $G_1'$,
$(v,u)$ precedes $(v,w)$ in the clockwise order of edges around
$v$ in $G_1$ if and only if
$(f(v),f(u))$ precedes $(f(v),f(w))$ in 
the clockwise order of edges around $f(v)$ in $G_2$.
That is, we
match vertices consistently with the edge orderings around each vertex.
\end{enumerate}
Since road evolution tends to involve adding or removing whole roads or 
neighborhoods, we restrict our notion of road network evolution in this paper
to be defined in terms conformal matchings.
There is still one more restriction that we need to add, however, which deals
with degeneracies that are unlikely to occur in real-world road networks.

Suppose we are
given two road networks, $G_1$ and $G_2$, and
a maximum-cardinality subgraph, $G_1'$, having 
a conformal matching, $f$, to a subgraph of $G_2$ (which is how we determine
the parts of $G_1$ that are the same in $G_2$).
We say that $G_1$ and $G_2$ are \emph{degenerate} 
if, for any vertex $v$ in $G_1'$ and edge $(v,w)$ in $G_1$, 
we can change the assignment, $f(w)$, for $w$
and still have $f$ be a conformal matching 
(even allowing for $f(w)$ to be undefined).
Since our intended applications involve the second road network being
a newer copy of the first, a maximum-cardinality subgraph with a conformal
matching identifies
the portions of the road network that have not changed over time; hence,
the portions outside of this maximum-cardinality subgraph identify the
portions that have changed.
Thus, we argue that such applications involve non-degenerate graphs, since
it is unlikely, for example, for us to encounter
an $8\times 16$ grid that evolves into a grid-like annulus of 64 nodes
with radius 8, which would be degenerate.
Given that such configurations are likely to be rare in the real world,
we are interested in this paper only in 
finding maximum conformal matchings in 
non-degenerate pairs of road networks, which is a problem
we refer to as the \emph{map evolution} problem.

Incidentally, 
the map evolution problem should not be confused with the \emph{map matching}
\ifFull
\else
problem (e.g., see~\cite{Liu:2012,Lou:2009}),
\fi
which is the unrelated problem of matching a trajectory of 
(possibly noisy) GPS
coordinates, as might be produced by a moving vehicle, to the
geometry of the road network in which the trajectory is traveling.


\subsection{Prior Related Work}
As noted above, the map evolution
problem is related to the graph isomorphism problem,
which has a rich history (e.g., see~\cite{grohe2000isomorphism,McKay201494}), 
due to the fact that it is not known
to be NP-complete, and the best known worst-case algorithm
runs in quasipolynomial time~\cite{babai2015graph},
but the problem tends to be feasible
in practice (e.g., see~\cite{McKay201494}).
For the generalized approximate graph isomorphism problem,
which is NP-hard, Arvind {\it et al.}~\cite{Arvind2012} give a 
quasipolynomial approximation algorithm.
Such algorithms are necessarily not taking advantage of any efficiencies,
however, that could come from topological considerations like our restrictions
to embedded graphs and conformal matchings.


The map evolution problem is also related to 
the \emph{map alignment} problem,
which is also known as \emph{GIS conflation}
\ifFull
\else
(e.g., see~\cite{rosen1985match,saalfeld1988conflation}).
\fi
In this problem, one is given two road networks, including both topological
information (such as vertex-edge-face relationships)
and geometric information (such as vertex coordinates and edge directions
and lengths),
and one is interested in computing a ``most likely'' matching between
the two networks.
Rosen and Saalfeld~\cite{rosen1985match,saalfeld1988conflation}
develop an iterative process involving a human operator
based on matchings that use
topology/geometry classifications of the vertices,
edges, and faces of the maps.
Xiong~\cite{xiong2000three} extends these topological/geometric
approaches using more sophisticated classifications.
Savary and Zeitouni~\cite{savary2005automated} 
and
Zhang~\cite{zhang2009methods} extends these approaches further by including 
additional properties, such as geographic data, including road names and shapes.
Their use of geometry, however, implies that all of these conflation 
methods are not strictly topological algorithms and their performance degrades
when roads or vertices lack geometric or geographic information.

Detecting changes in road networks 
and geographic regions has also been studied from the perspective
of image processing, e.g., using satellite images
\ifFull
\else
(e.g., see~\cite{ventura1990image}).
\fi
For example,
Zhang and Couloigner~\cite{qiaoping2004automatic} use image analysis
to extract polylines defining roads and match them between two images
of the same geographic region taken at different times.
Such image-analysis approaches are inherently geometric, however; hence these
are also not strictly topological algorithms and do not apply when image 
data is not available.

Our topological approach is more closely aligned with the work
of Eppstein {\it et al.}~\cite{eppstein2009approximate}, which uses
a topological approach for approximately
matching for quadrilateral meshes used in 
computer-generated animations. 
Our approach differs from their methods, however,
in that we do not consider faces in our matching algorithm (since road
network faces can be large and complex), whereas their method crucially depends
on matching faces (which in their application are always
quadrilaterals or triangles).

\subsection{Our Results}
In this paper, we study the map evolution problem,
for matching two road network
graph of same area but from different time, by using only topological
properties. 
The primary motivation for this approach is to show that the map 
evolution problem
problem can be solved effectively using only topological information.
Thus, this gives GIS practitioners a tool that can be applied for solving
the map evolution problem even for problem instances where geometric and
geographic information is missing, such as in older hand-drawn maps, pairs of
maps where only one of them is derived from an image, pairs of 
maps annotated in different languages, or maps missing geographic and
geometric annotations due to scaling resolution.

We develop an algorithm for the map evolution
problem that runs in polynomial time for finding conformal matchings
between non-degenerate embedded graphs, such as real-world road networks.
Our algorithm uses a breadth-first flooding technique that begins each
flooding phase by finding potentially-matching ``seed'' vertices using
a labeling technique similar to that used in the
the Weisfeiler-Leman (WL) graph isomorphism
algorithm (e.g., see~\cite{grohe2000isomorphism}). 
So as to limit the amount of flooding done in subgraphs that ultimately are
determined not to match,
our algorithm is probabilistic in nature---when we don't have any
pair of unique starting nodes, we take the pair which minimizes 
an estimate of the probability of a wrong match. 
We provide verification of our algorithm
in experiments 
and case studies 
that show empirically that our algorithm produces good matches
in practice.


\section{Our Algorithm}
In this section, we describe our topological algorithm for
finding a best conformal matching between 
two non-degenerate road networks, $G_1$ and $G_2$:
\begin{enumerate}
\item
Create quasi-unique labels for each vertex, $v$,
in $G_1$ and $G_2$ based on the degrees of the nodes 
at distance at most $k$ from $v$, for an input parameter, $k$.
(We show in our experimental section that choosing $k$ 
between 5 and 8 tends to give the best results.)
\item
Choose a good 
pair of starting nodes, $s_1 \in G_1$ and $s_2\in G_2$, with the same
quasi-unique label, $L$, and, for each such pair having label $L$, 
perform the following:
\begin{enumerate}
\item
Perform a breadth-first search (BFS) matching of the corresponding portions 
in $G_1$ and $G_2$ that are respectively reachable from $s_1$ and $s_2$
according to a greedy conformal matching that emanates out from these starting
nodes.
\item
Save this conformal matching that starts from $s_1$ and $s_2$ if it is the best 
(highest cardinality) such matching found so far for this quasi-unique label.
\end{enumerate}
\item
Commit the conformal matching that began with $s_1$ and $s_2$, removing
all matched nodes as candidates for starting nodes.
\item 
Repeat the above process for another good pair of starting nodes, if such
a pair of nodes still remains.
\end{enumerate}

We describe these steps in more detail below.

\subsection{Labeling Vertices}
The first step of our algorithm is 
to give each vertex, $v$, in $G_1$ and $G_2$ a quasi-unique label, based on 
the degrees of the nodes at distance at most $k$ from $v$, for a given
parameter, $k$.
This approach is similar to a labeling method used in
the (exact) graph isomorphism 
algorithm by Weisfeiler and Leman (WL)~\cite{grohe2000isomorphism}.
Specifically, we begin by determining the degree, deg$(v)$, of each 
vertex, $v$.
Then we create a list for each vertex, $v$,
which contains its degree, followed by the degrees of nodes at distance $1$
from $v$, nodes at distance $2$ from $v$, and so on, up to a distance $k$,
where $k$ is an input parameter for this step.
So as to make sure that these labels are quasi-unique, we add the degrees of
these nodes at distance at most $k$ from $v$ according to a canonical
ordering, which in our case is a lexicographically minimum breadth-first
search (BFS) ordering. 
This BFS ordering sorts the immediate neighbors of $v$ according
to a lexicographically minimum cyclic ordering of $v$'s neighbors based on
their degrees, and then it performs a 
BFS from this queue, adding nodes
to the queue based on the cyclic ordering of edges around each vertex so long
as they are at distance at most $k$ from $v$.

%

We return a dictionary for $G_i$ (for $i=1,2$),
which we call {masterTable}$({G_i})$, such 
that each entry in this dictionary is a list of vertices having the same
quasi-unique label.
That is, the keys we use to index the (list) entries in masterTable$({G_i})$
are the label$[v]$ lists produced by our quasi-labeling method.

The pseudocode for this step is given in Algorithm \ref{alg:find_seed}. 

\begin{algorithm}
    \SetKwInOut{Input}{Input}
    \SetKwInOut{Output}{Output}
\medskip
    {\textbf{function} labelNodes}$(k,G)$\;
	    \For{each $v \in G$}{
	          label[$v$] = (deg$(v)$) \hspace*{2em} \# label is a list\;
	        \For{$u \in \mbox{lexicographicBFS}(v,k)$}{
	              Append deg($u$) to end of label[$v$]\;
	        }
	          Add $v$ to masterTable$(G)\,$[label$[v]]$\;
		}
    \caption{Algorithm for labeling each vertex with a quasi-unique label.
     The method, lexicographicBFS$(v,k)$, 
     returns an ordered list of nodes as would be visited
     a breadth-first search (BFS) from $v$, starting with the neighbors
     of $v$ enqueued according to a lexicographically minimum cyclic 
     ordering of $v$'s neighbors based on their degrees. This BFS
     explores all nodes at distance at most $k$ from $v$.}
    \label{alg:find_seed}
\end{algorithm}

With respect to the efficiency for performing this step, note that 
the time needed for this step is dominated by our doing
a BFS from each node, $v$, to explore those other nodes at distance $k$
from $v$.
Since the vertices of a road network have degree bounded by some parameter,
$d$, this step takes worst-case time $O(d^kn)$, for a road network of $n$ nodes.
In practice, $k$ is a constant, $d$ is usually $3$ or $4$,
and the graph is rather sparse; hence, 
this step runs in $O(n)$ time in practice.

\subsection{Choosing Pairs of Starting Nodes}
After we have labeled each vertex of $G_1$ and $G_2$ with quasi-unique
labels, we need to choose a pair
of starting nodes in $G_1$ and $G_2$ with the same 
label to start the matching process. 
If we are able to
find a unique pair of nodes having the same label, then we can take them
as starting nodes and start our matching. 
But it may happen that
we don't have any such unique pair of nodes; that is,
it might be the case that
there are at least $3$ nodes from $G_1\cup G_2$
for each quasi-unique label of vertices in the master table.

For each distinct label, $L$, 
let $n_1(L)$ denote the number of vertices in $G_1$ with label $L$
and let $n_2(L)$ denote the number of vertices in $G_2$ with label $L$.
As mentioned above, if we have a label, $L$, such that
$n_1(L)=n_2(L)=1$, then we choose the unique pair of vertices,
$s_1\in G_1$ and $s_2\in G_2$, with label $L$ 
as a good pair of starting vertices.

Otherwise, we would like to choose a pair,
$s_1\in G_1$ and $s_2\in G_2$, that maximizes the probability that 
there is a large conformal matching of the connected components
of $G_1$ and $G_2$ respectively containing $s_1$ and $s_2$, such that
$s_1$ and $s_2$ have the same quasi-unique label, $L$.
For any such label, $L$, the number of such candidate pairs is 
$n_1(L)\cdot n_2(L)$; hence, to maximize the probability of finding a good
pair of starting nodes, we choose a pair, $s_1$ and $s_2$, that minimizes
the product, $n_1(L)\cdot n_2(L)$,
since the probability such 
a pair actually correspond to corresponding nodes in $G_1$ and $G_2$, 
conditioned on their having the same label, $L$,
is at least $1/(n_1(L)\cdot n_2(L))$.

We then perform a flooding-based search from each such $s_1$ and $s_2$ with
label $L$, committing to the pairing that results in the largest matched
components in $G_1$ and $G_2$.
Then, we remove all the matched vertices
in $G_1$ and $G_2$ from consideration (since they are now matched), 
and we repeat our search for another good pair of starting seed vertices.

In order to perform such searches and updates quickly, we use an auxiliary
priority queue
data structure that stores each quasi-unique label, $L$, according
to its priority, $n_1(L)\cdot n_2(L)$.
Such
products can be found by taking the product
of lengths of both lists for each label used as a key
in \emph{masterTable}. 
As we are performing our greedy matching processes,
we also need to
update these lists by removing each matched pair of nodes.
Of course,
this will also change the product for each label, so we have to update
labels in our priority queue to now have new priorities.
Since these products are always integers in the range $[1,C]$,
for some parameter, $C\le n^2$, let us use
a van~Emde~Boas tree~\cite{VANEMDEBOAS197780,vanEmdeBoas1976}
(vebTree) for storing non-zero products,
$n_1(L)\cdot n_2(L)$, for each label, $L$, as well as a
hash table, \emph{productTable}, that gives us the product for any existing
label, $L$.
This allows us to perform searches, updates, and finding of labels with
minimal product values in $O(\log\log C)$ time.

Every time the algorithm needs a pair of starting nodes, it finds
a label, $L$, with minimum 
product, $n_1(L)\cdot n_2(L)$,
from \emph{vebTree}.
If there are multiple labels having that product, we randomly choose
any one of them. 
After finding the required label, we take a pair
of nodes having the same label from the \emph{masterTable}. After
finding the starting pair of nodes, we update these data structures,
and the productTable,
so that we don't consider this pair of nodes again.
Note
this approach works even when we have unique pair of nodes having
the same label. In that scenario, the product will be $1$ and that
will be minimum product in \emph{vebTree}.

With respect to efficiency, we can do all the setup for this step
in $O(n\log\log C)$ time.
Moreover, we can determine already at this point what is the maximum
product, $n_1(L)\cdot n_2(L)$, over all labels, $L$, for a given
value of $k$. Since $k$ is a constant for real-world road networks
and there is an inverse relationship between $k$ and the size of
these products, we can perform a (binary) search to choose $k$ 
so that the maximum product size is bounded by some constant, $C$.
The running time of this search would be $O(n)$ for constants $k$ and
$C$.

There is a tradeoff, however, between using a large value for $k$ and getting
good matches, since two starting nodes are paired only if their
quasi-unique labels are the same, that is, if the
respective portions of the road
network at distance $k$ from these nodes is the same. 
Since we are considering road networks that are evolving, we 
therefore don't want to set too high a value for $k$.
Thus, we would like to choose $k$ as small as possible so that the
products, $n_1(L)\cdot n_2(L)$, are bounded by a constant, $C$.
(Say, $C\le 24$.)
As we note in the experimental section of this paper, choosing $k$
between 5 and 8 seems to work well in practice for this purpose.

\subsection{Flood-based Conformal Matching}
After finding a starting pair of nodes, we start our 
greedy BFS matching process.
We begin by marking the starting nodes as
matched and we add them to our current tentative matching.
As we perform our BFS matching process, we will tentatively be matching up
additional pairs of nodes from $G_1$ and $G_2$, updating our
supporting data structures as we go, e.g., to tentatively remove each 
such pair from consideration in vebTree.
Moreover, 
if a starting node has more
than one lexicographically minimum ordering of the degrees of its
neighbors, then we also consider each such ordering of the edges, 
performing our BFS matching process for each.
Tentative matchings are compared on the basis of number of matched nodes
and the matching with maximum number of matched nodes is taken as
best matching.

This raises an important implementation detail, which we should
probably discuss before going on to other details.  Our matching algorithm
considers different pairs of starting vertices (and even 
possibly different starting orientations of their incident edges), looking for
the pair that produces the largest portions of matching subgraphs.
Thus, we may have tentative matches that need to be undone
so that other tentative matches can be considered. 

There are at least two possible ways to deal with this
branch-and-bound element in our conformal matching 
algorithm.
One way is to checkpoint our supporting data structures, like
vebTree, masterTable, and productTable, saving the 
version that produced the best tentative match so far.
This is the method we use, for example, in the version 
of our algorithm that we implemented
for our experiments, since it is easy to implement.
Another way is to perform a two-phase commit, where we perform updates
to global copies of these data structures, but keep a history of the
updates we have performed during a tentative matching, so that we can
then roll back these updates if we do not commit to that tentative
matching (because there is another one that gave a larger number of
matched vertices).
This is the version of our algorithm
that we analyze for our theoretical analysis.

Given that there is some method that allows us to roll back to an
earlier state of our supporting data structures, vebTree,
masterTable,
and productTable, let us discuss in more detail how our conformal BFS
proceeds.

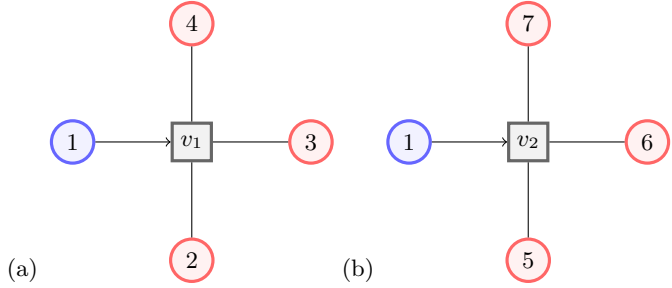
\begin{figure}[hbt!]
\sidesubfloat[]{%
    \begin{tikzpicture}[
        roundnode/.style={circle, draw=red!60, fill=red!5, very thick, minimum size=3mm},
        squarednode/.style={rectangle, draw=black!60, fill=black!5, very thick, minimum size=5mm},
        donenode/.style={circle, draw=blue!60, fill=blue!5, very thick, minimum size=3mm},
        ]
        \node[squarednode]  (maintopic)                              {$v_1$};
        \node[donenode]    (leftcircle)        [left=of maintopic]  {1};
        \node[roundnode]    (lowercircle)       [below=of maintopic] {2};
        \node[roundnode]    (rightsquare)         [right=of maintopic] {3};
        \node[roundnode]    (uppercircle)       [above=of maintopic] {4};
        
        \draw[->] (leftcircle.east) -- (maintopic.west);
        \draw[-] (maintopic.south) -- (lowercircle.north);
        \draw[-] (maintopic.east) -- (rightsquare.west);
        \draw[-] (maintopic.north) -- (uppercircle.south);
    \end{tikzpicture}
}
\sidesubfloat[]{%
    \begin{tikzpicture}[
        roundnode/.style={circle, draw=red!60, fill=red!5, very thick, minimum size=3mm},
        squarednode/.style={rectangle, draw=black!60, fill=black!5, very thick, minimum size=5mm},
        donenode/.style={circle, draw=blue!60, fill=blue!5, very thick, minimum size=3mm},
        ]
        \node[squarednode]  (maintopic)                              {$v_2$};
        \node[donenode]    (leftcircle)        [left=of maintopic]  {1};
        \node[roundnode]    (lowercircle)       [below=of maintopic] {5};
        \node[roundnode]    (rightsquare)         [right=of maintopic] {6};
        \node[roundnode]    (uppercircle)       [above=of maintopic] {7};
        
        \draw[->] (leftcircle.east) -- (maintopic.west);
        \draw[-] (maintopic.south) -- (lowercircle.north);
        \draw[-] (maintopic.east) -- (rightsquare.west);
        \draw[-] (maintopic.north) -- (uppercircle.south);
    \end{tikzpicture}
}
\caption{Neighbor ordering around a degree-$4$ node, $v_1$, and its
matching node, $v_2$.
In this example, node 1 in $G_1$ matches with node 1 in $G_2$, and
we know that the we reached the matched nodes $v_1$ and $v_2$
through the respective nodes, $1$, so now their
clockwise ordering is fixed and the mapping of neighbors will be
$(4,7), (3,6), (2,5)$.
}
\label{fig:order}
\end{figure}

Once we map the neighbors around a pair of starting nodes, as discussed
above, we flood-search both graphs using a conformal-matching BFS. 
When we reach any other
node except a starting node in the flooding, we know the edge we
are coming from and as we are following clocking ordering around
any node, there will be exactly one ordering around that node in
which we can traverse and map the neighbors with another graph, so as
to be forming a conformal matching.
Figure \ref{fig:order} shows an example.

For matching any two nodes, $v_1\in G_1$ and $v_2\in G_2$, 
that are not starting nodes, they should satisfy following properties:
\begin{itemize}
    \item both $v_1$ and $v_2$ should be unmatched.
    \item degree of $v_1$ should be same as $v_2$.
\end{itemize}
If any of these two conditions fail, we don't match $v_1$ and $v_2$
and we terminate that branch of the BFS. If both the conditions are satisfied,
then we mark $v_1$ and $v_2$ as matched, add them to current matching
and the queue for the BFS.
Then we remove them from \emph{masterTable}, \emph{vebTree} and
\emph{productTable}, so that they are not considered again in 
the matching
process. The pseudocode for this step in our algorithm is given as Algorithm
\ref{alg:process_nodes}.


\begin{algorithm}
    \SetKwInOut{Input}{Input}
    \SetKwInOut{Output}{Output}
    
    {\textbf{function} processNodes}($u_1,u_2$)\;
	    \Indp add ($u_1$, $u_2$) to matching\;
        mark $u_1$ and $u_2$ as matched\;
        add corresponding neighbors of $u_1$ and $u_2$ to bfsQueue\;
        update masterTable, vebTree and productTable to remove $u_1$ and $u_2$\;
    \caption{Algorithm to process nodes in BFS.}
    \label{alg:process_nodes}
\end{algorithm}

When there is no further branch that can be matched, our BFS search terminates.
If this is the best tentative matching for the given quasi-unique
label, $L$, 
then we tentatively save the matching corresponding to this BFS to the total
matching. 
Then we check if there is still any remaining pair of seed nodes
having this same label. 
If so, then we perform another conformal BFS for this next pair of
seed vertices.
Once we have completed performing a tentative matching for each pair
of seed nodes having the same quasi-unique label, $L$, we commit to
the matching for this label that produced the largest match.

Then we check if
\emph{vebTree} is empty or not.
If \emph{vebTree} is empty, we terminate the
algorithm and return the total matching. If not, we repeat our search
for a quasi-unique label, $L$, having the smallest product, 
$n_1(L)\cdot n_2(L)$, and repeat the above conformal BFS for that
label.

The pseudocode for this step in
our algorithm is given as Algorithm \ref{alg:matching}.


\begin{algorithm}[bht]
    \SetKwInOut{Input}{Input}
    \SetKwInOut{Output}{Output}

    {\textbf{function} matching}(masterTable,$G_1,G_2$)\;
         create prodTable and vebTree\;
	     totalMatching = [ ]\;
	    \While{vebTree is not empty}{
	         minProd = vebTree.min$()$\;
	         startingLabel = productTable[minProd]\;
	         startingPairs = 
		 (masterTable$(G_1)\,$[startingLabel], 
		                  masterTable$(G_2)\,$[startingLabel])\;
	         find all mappings of neighbors around each pair in 
		 startingPairs\;
	        \For{each of the mappings in a startingPair}{
	             matching = [ ]\;
	             $(s_1,s_2)$ = this instance of startingPair[0],startingPair[1]\;
	             bfsQueue = $()$\;
	             processNodes($s_1$,$s_2$)\;
	            \While{bfsQueue is not empty}{
	                 $v_1, v_2$ = pop(bfsQueue)\;
	                 \If{$v_1$ and $v_2$ are both unmatched}
	                 {
	                    \If{deg$(v_1)$ = $deg(v_2)$}
    	                {
    	                     processNodes($v_1$,$v_2$)\;
    	                }
	                 }
	            }
	         checkpoint this matching if it's best 
		 for this startingPair\;
	        }
		 add the best matching found to totalMatching\;
	    }
    \caption{Our flood-based conformal matching algorithm.}
    \label{alg:matching}
\end{algorithm}

Each time
we explore subgraphs of $G_1$ and $G_2$ for a particular starting
pair, $s_1$ and $s_2$, that are in the starting label set of pairs
for some quasi-unique label, $L$, and one of the deg$(s_1)$ possible
orientations of edges, we traverse subgraphs of some size at most,
$n(L)\le n$, where $n(L)$ is the size of the largest match for the
label $L$.
Thus, the running time of this part of our algorithm is 
at most
$O(n(L)\log\log C)$, where $C$ is the maximum value of a
product, $n_1(L')\cdot n_2(L')$, for some label, $L'$.
If $d$ is the maximum degree in a road network (e.g., $d\le 8$), 
then the total worst-case running time of our BFS matching algorithm is
therefore
\[
O\left(dC\sum_L n(L)\log\log C\right)
=
O(dCn\log\log C),
\]
since $\sum_L n(L) \le n$, because the maximum amount of nodes we can
ultimately match in a pair of non-degenerate road networks is $n$.
Combining this with the theoretical analysis of the other steps
in our matching algorithm implies that the total running time
of our entire algorithm is
$O(d^kn+dCn\log\log C)$, where $d$ is the maximum degree of
a road network, $k$ is the distance we choose for producing
quasi-unique labels, and $C$ is the maximum value of a product,
$n_1(L)\cdot n_2(L)$, for any label, $L$.
Thus, in the practical case when $d$, $k$, and $C$ are constants, our
matching algorithm runs in $O(n)$ time.


\begin{figure*}[hbt!]
\begin{center}
    \subfloat[{$k = 1$}]
        {
                \label{fig:amador-hist-1}
                \includegraphics[width=0.28\textwidth]{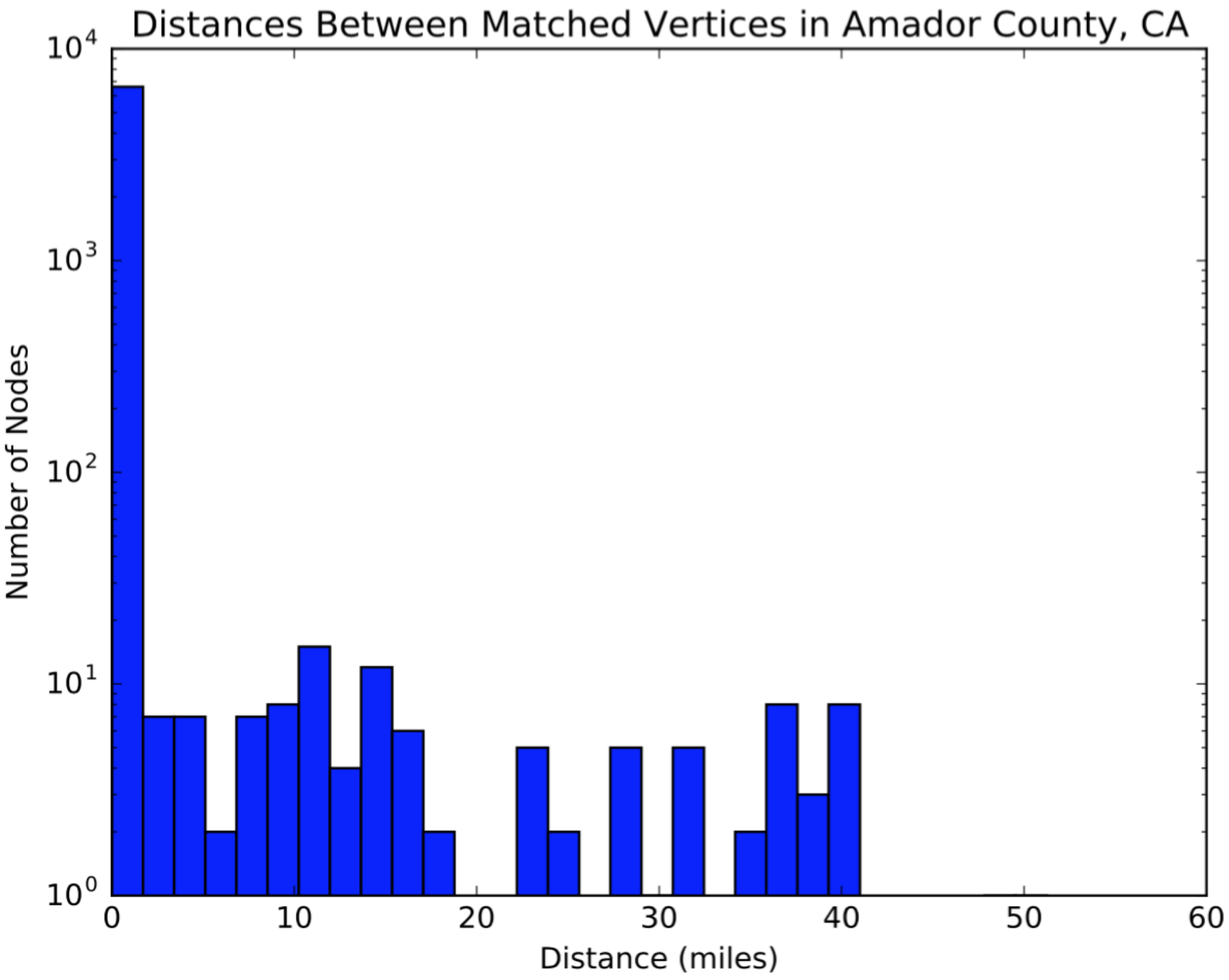}
        }
	\subfloat[{$k = 5$}]
        {
                \label{fig:amador-hist-2}
                \includegraphics[width=0.28\textwidth]{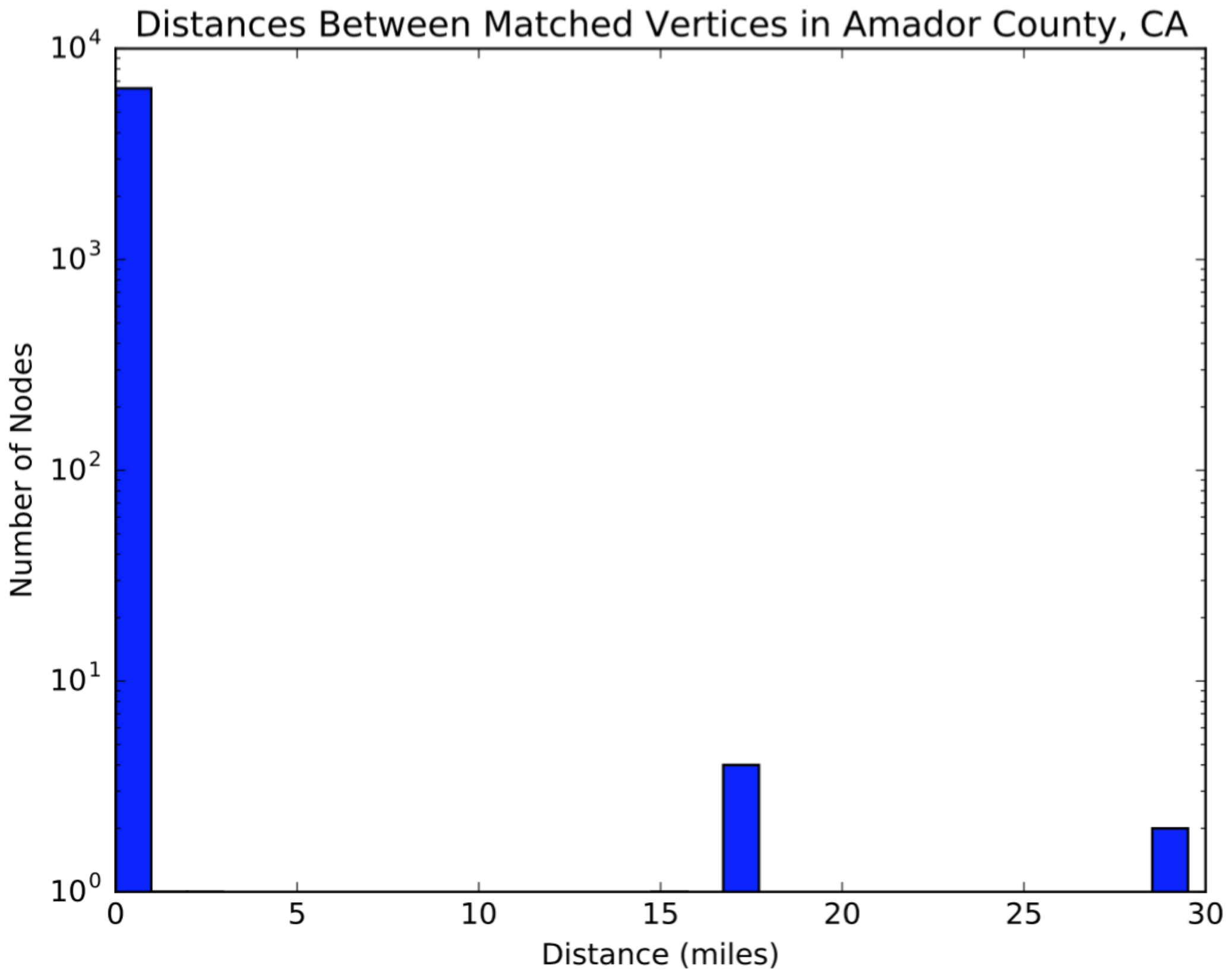}
        }
\end{center}
\vspace*{-18pt}
\caption{Histogram plots for Amador County, CA from 2000 to 2006.}
\label{fig:amador-hists}
\end{figure*}

\begin{figure*}[hbt!]
\begin{center}
    \subfloat[{$k = 1$}]
        {
                \label{fig:sc-hist-1}
                \includegraphics[width=0.28\textwidth]{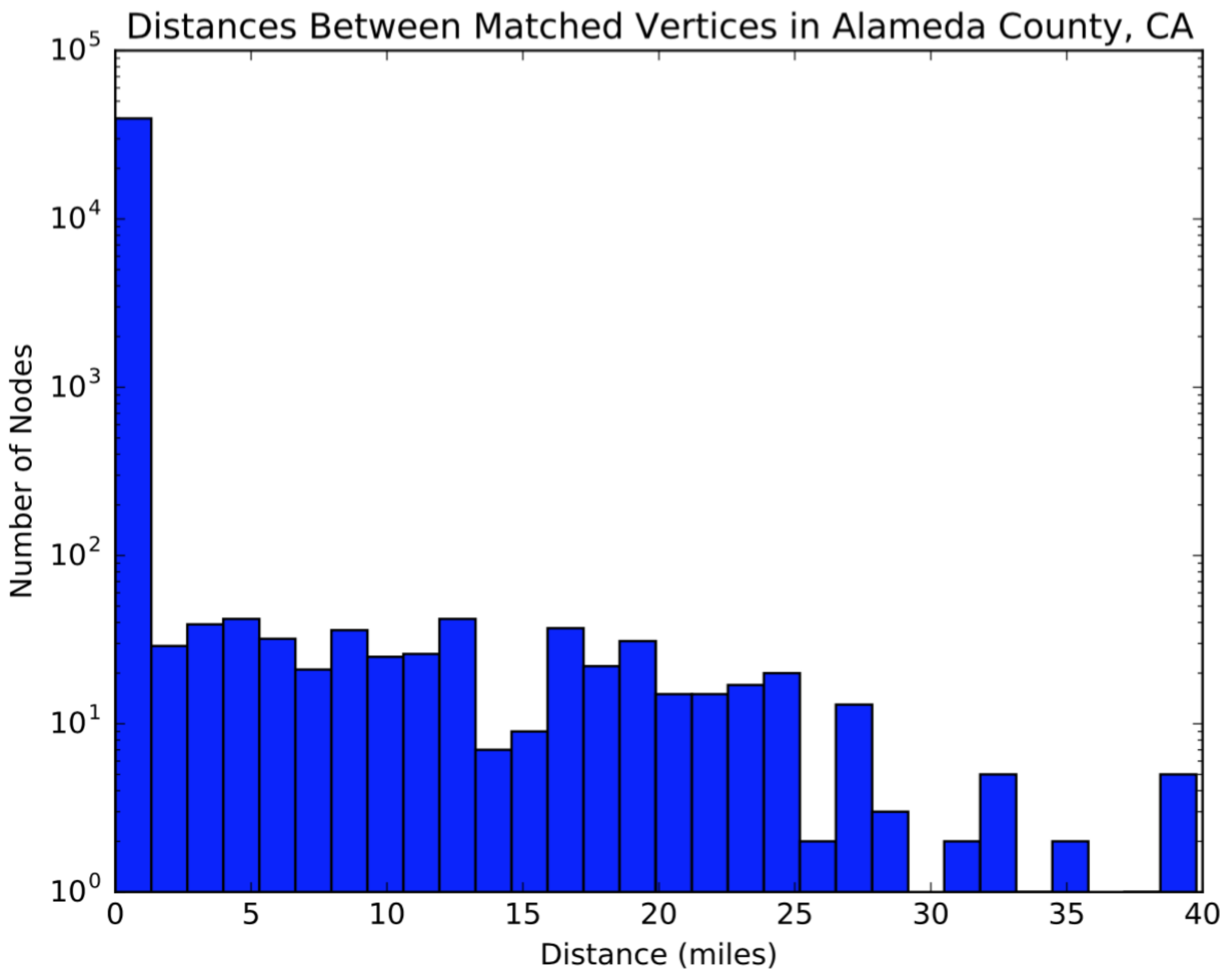}
        }
	\subfloat[{$k = 7$}]
        {
                \label{fig:sc-hist-5}
                \includegraphics[width=0.28\textwidth]{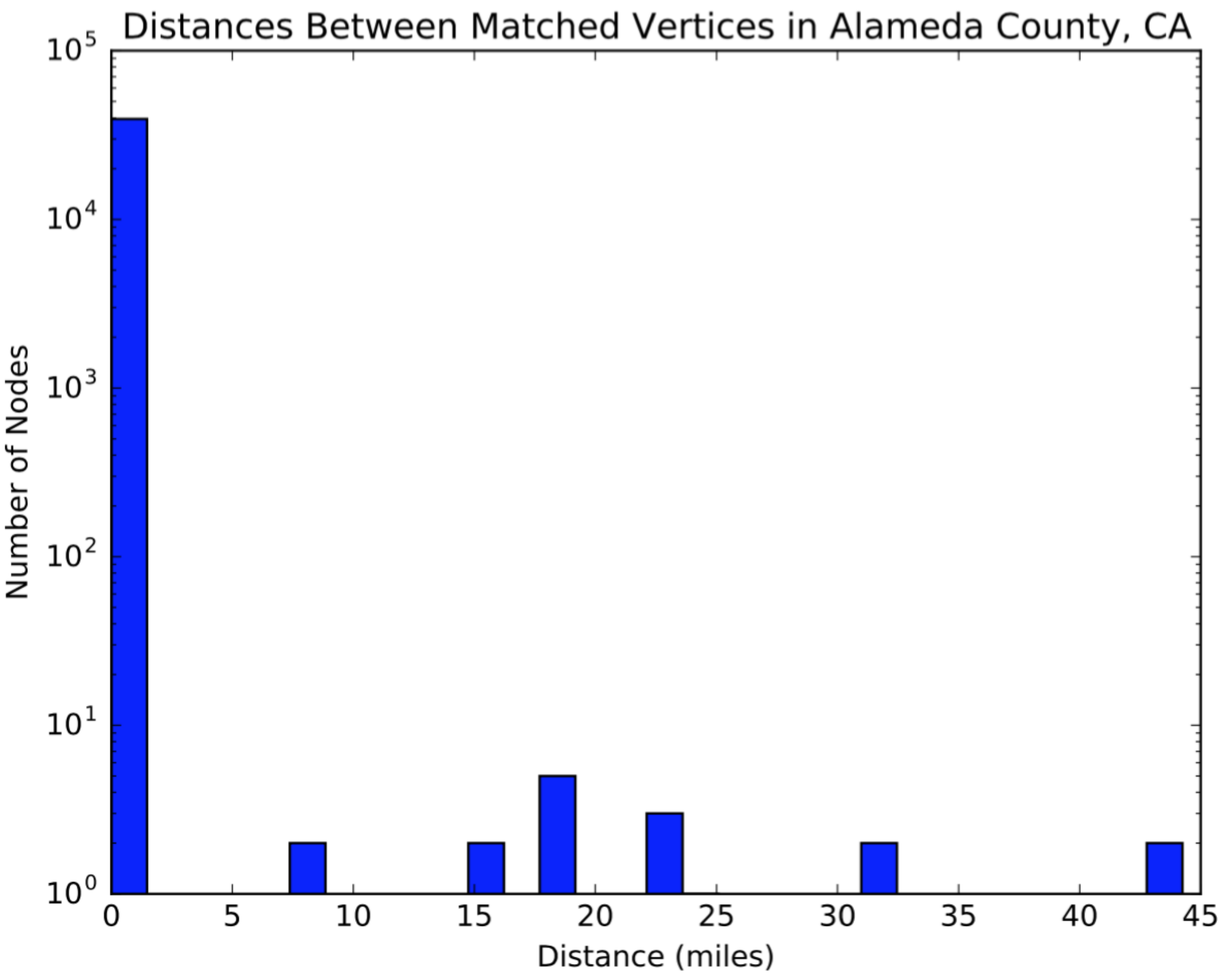}
        }
\end{center}
\vspace*{-18pt}
\caption{Histogram plots for Alameda County, CA from 2000 to 2006.}
\label{fig:alameda-hists}
\end{figure*}

\section{Experiments}

In this section, we provide an empirical evaluation of our topological flood-based matching.  All of our experiments were ran on data from the U.S. TIGER/Line road network database~\cite{tiger-line-database}.

\subsection{Preprocessing the Data}

The TIGER/Line database provides the road networks in two different file formats: shapefile and TIGER/Line ASCII format.  The data the shapefile format provides allows a graph to be created that not only has a node for every intersection of two roads, but also nodes to indicate the curvature of a road.  That is, the format allows for curved roads to be represented as a sequence of many two-degree vertices.  Therefore, for the preprocessing of files in the shapefile format, we simply take the first and the last vertex for each road to avoid introducing unnecessary two-degree vertices.  With this approach to processing files in the shapefile format and fact that the TIGER/Line ASCII format lends itself to easy conversion to the definition of a road network given in the introduction, our algorithm performs well on both file formats.

\subsection{Tuning the Seed-labeling Parameter}

Let us consider the choice of the value for $k$, the parameter that 
is input to Algorithm~\ref{alg:find_seed} 
that defines distance to which to perform a lexicographic BFS so
as to improve the uniqueness
of vertex labels.  
To characterize this uniqueness factor, let us
define the \emph{approximation ratio} of a
labeling as $a/b$, where $a$ is the number of pairs of nodes 
with the same label
and $b$ is the minimum of the number of nodes in the two graphs.
Intuitively,
if $k$ is small, there will likely be many pairs of nodes $(u,v)$ with $u$
in $G_1$ and $v$ in $G_2$ that both have label $L$ where 
$n_1(L) \cdot n_2(L)$ is large.  
For example, labels
like ``44444", which indicates a four-way intersection that leads
to four other four-way intersections, are likely to
be common, and many other
examples like this are likely from real-world.  As many of these
products are expected to be large, 
we would expect the approximation ratio to be larger for
smaller value for $k$, because we could be possibly finding many
pairs of vertices with the same label 
that should not actually be matched.  For instance, we might find
two vertices labeled ``44444" even though they are not similar
beyond their immediate neighbors.  We expect to run into this situation
only when the product is large since our algorithm matches the pair
of vertices for a given label that maximizes the number of nodes
matched.

As we increase the value of $k$, we would expect that the
approximation ratio to decrease.
That is,
if $k$ is large, we expect there to be more labels $L'$ such that $n_1(L')
\cdot n_2(L')$ is small or even 1, as the labels should become more
distinct as $k$ increases.  Because the labels are expected
to be more distinct in this case,
it should be less likely to find pairs of vertices with those labels,
causing the approximation ratio to decrease.

The histograms in Figures~\ref{fig:amador-hists} and
\ref{fig:alameda-hists} exemplify the preceding interpretation of
the parameter $k$.  The $x$-axis indicates the physical distance between
every node and its pair partner(s) with the same quasi-unique label,
$L$, using the longitude and latitude values given
from the database.  The distance is determined using the haversine
formula, which yields the shortest distance between two points on
a sphere~\cite{shumaker1984astronomical}.  
(Although our algorithm doesn't use geometric information to determine
matching pairs, we used geometric information in this experiment to
empirically validate our approach.)
Ideally, all pairs should be at distance 0 from each other.

As we expected, larger
$k$ values minimize the physical distances between pairs of nodes
with the same label, which
gives us a more accurate matching; hence, it reduces the number
of false pairs that our algorithm needs to consider.  A histogram that is highly
skewed is desirable, as that implies that the number of
incorrect nodes being falsely matched is small.  Note that the Amador County
data from 2000 and 2006 in Figure~\ref{fig:amador-hists} included
6,970 and 6,784 nodes, respectively, and the Alameda County data
from 2000 and 2006 in Figure~\ref{fig:alameda-hists} included 52,566
and 51,054 nodes, respectively.

\begin{figure}[htb]
\begin{center}
    {
        \includegraphics[width=0.7\columnwidth]{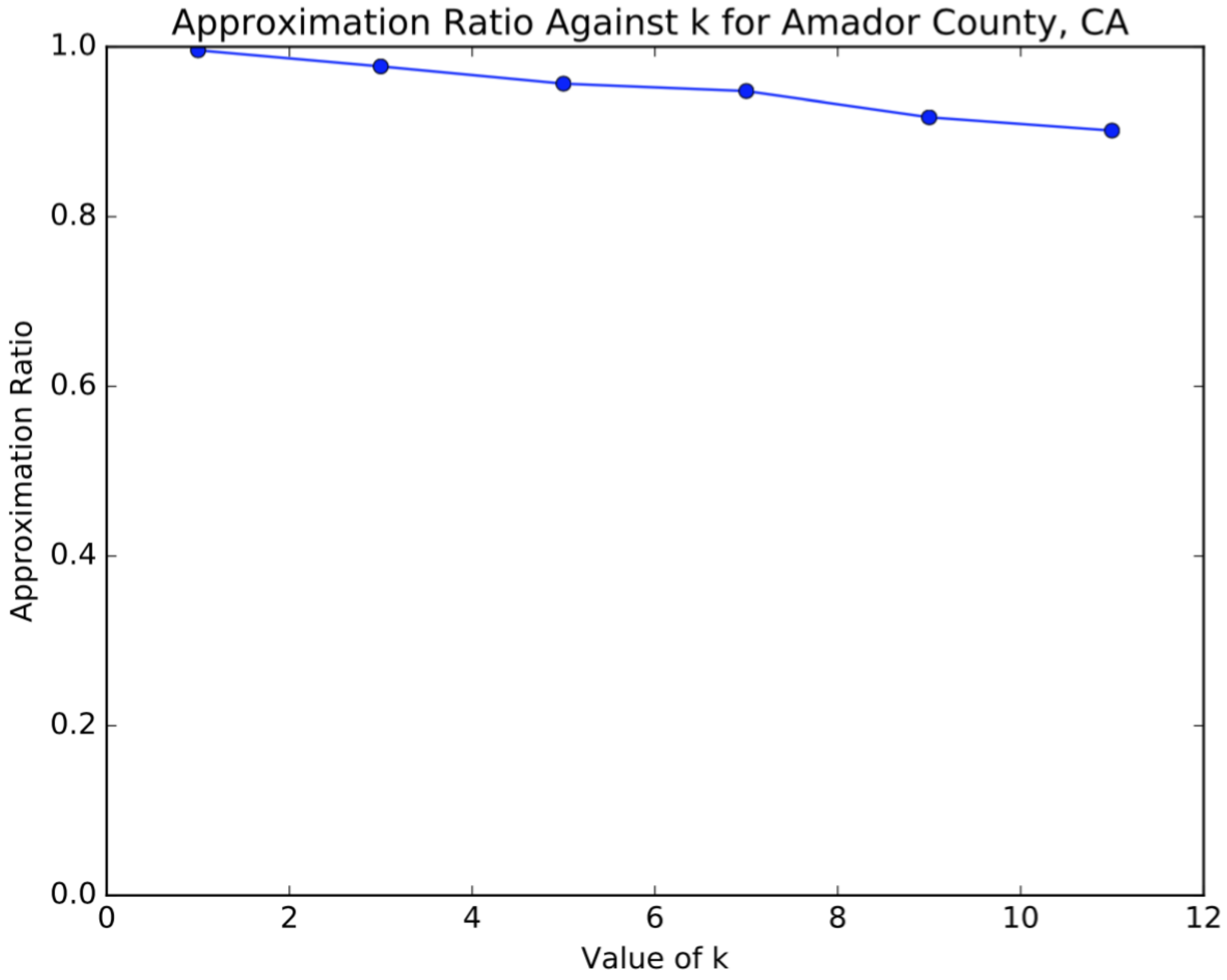}
    }
\end{center}
\vspace*{-12pt}
\caption{Change in approximation ratio for Amador County, CA from 2000 to 2006}
\label{fig:approx-plot}
\end{figure}

Figure~\ref{fig:approx-plot} shows the change of the approximation
ratio with respect to the change in $k$ for Amador County.  The
decrease in the approximation ratio with the increase in $k$ again
matched our intuition.  The plot with the same $x$-axis and $y$-axis
values for Alameda County started at a similar approximation ratio
and decreased at a similar rate, so it was omitted.

\begin{figure}
\begin{center}
    {
        \includegraphics[width=0.8\columnwidth]{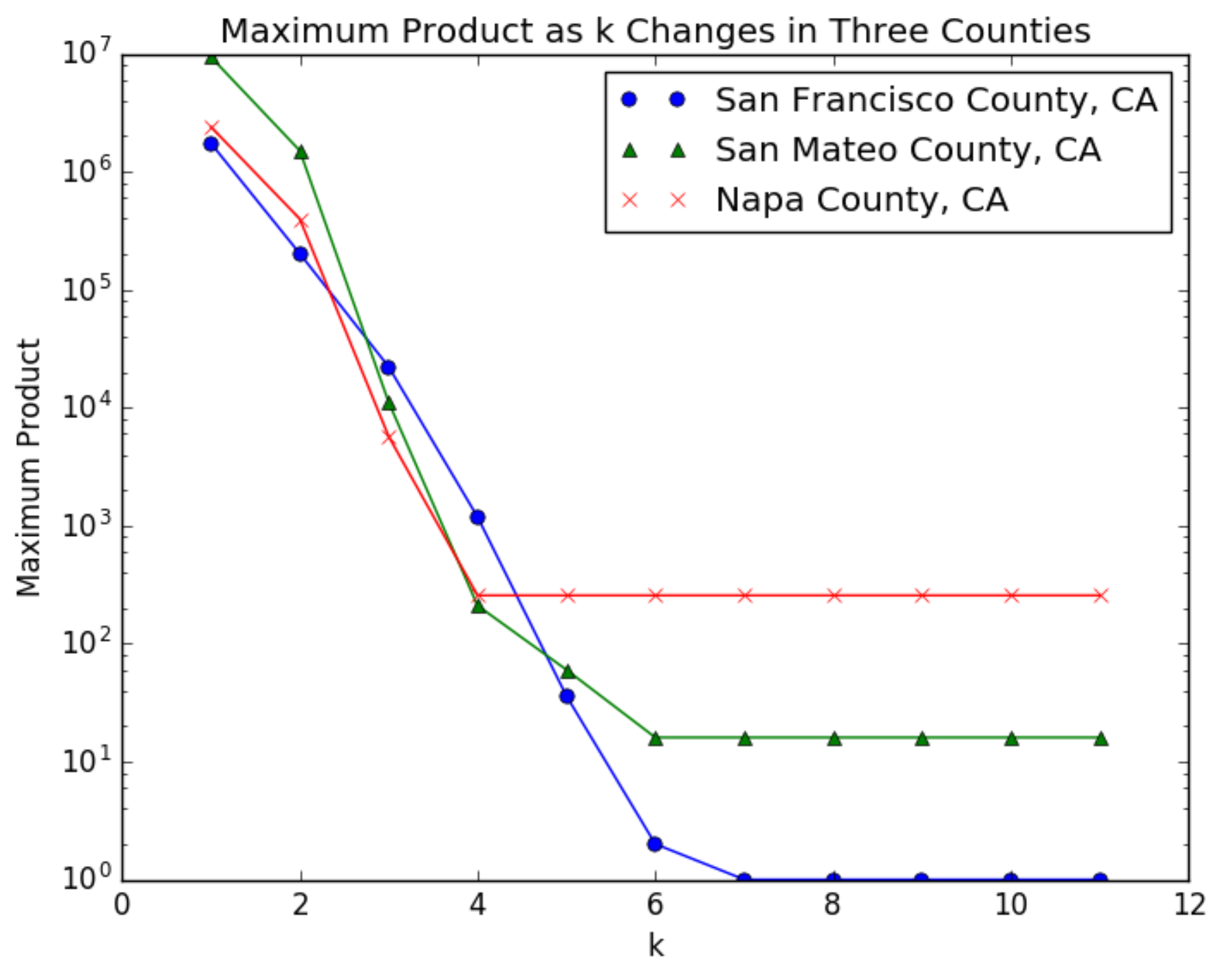}
    }
\end{center}
\vspace*{-12pt}
\caption{Change in maximum product for Napa, San Francisco, and San Mateo Counties with road networks from 2000 and 2006}
\label{fig:max-product}
\end{figure}

We also plot the change in the maximum product with respect to $k$ in Figure~\ref{fig:max-product}.  As described in Section 2, the maximum product is the value $\max\{n_1(L)\cdot n_2(L) : L \text{ is a label generated by Algorithm~\ref{alg:find_seed}}\}$.  As expected, the maximum product decreases as $k$ increases.  Note that only for San Francisco County does the maximum product reach 1.  This is due to the fact that for the other counties, there are labels that do not change as $k$ increases as the nodes the labels correspond to are in small connected components e.g.\ ``121" is the cause of this in San Mateo County.

\subsection{Example Output of Our Algorithm}

\begin{figure*}[hbtp]
\begin{center}
    \subfloat[{$2000$}]
        {
        	    \label{fig:ex-match-old-1}
                \includegraphics[width=0.3\textwidth]{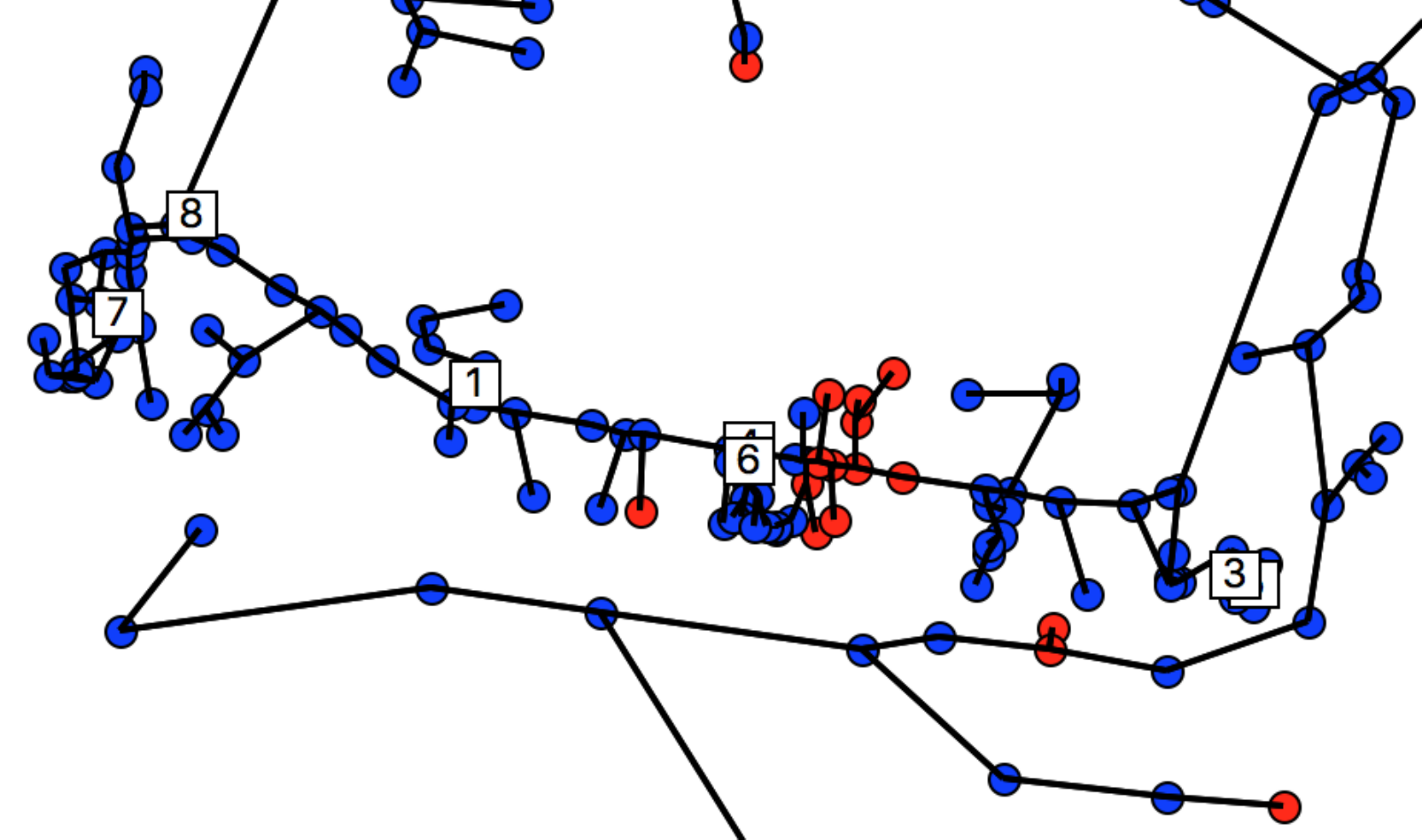}
        }
	\subfloat[{$2006$}]
        {
        	    \label{fig:ex-match-new-1}
                \includegraphics[width=0.3\textwidth]{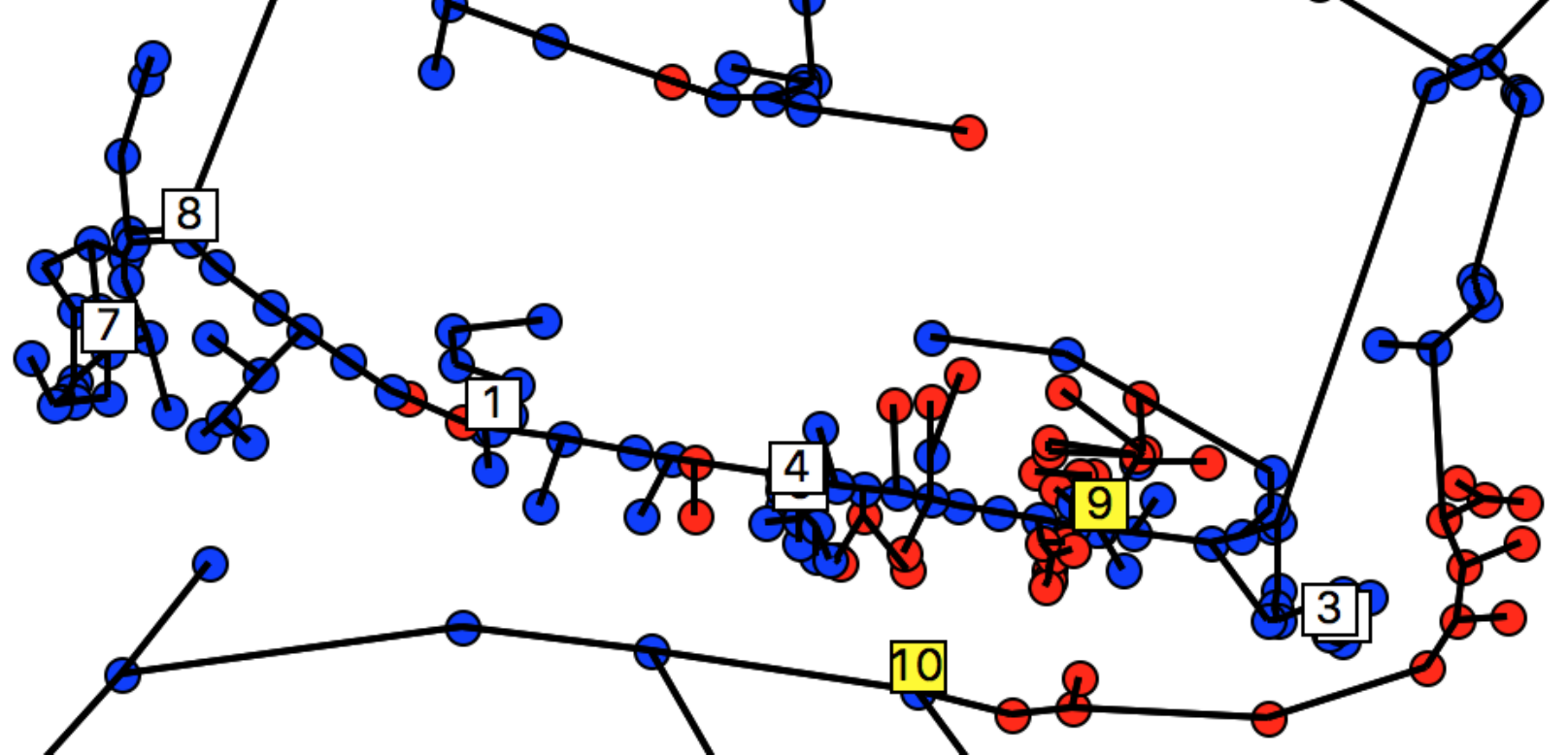}
        }
\end{center}
\vspace*{-12pt}
\caption{First example of a portion of a matching for Del Norte
County, CA where $k = 3$.}
\label{fig:ex-matching-1}
\end{figure*}

\begin{figure*}[hbtp]
\vspace*{-12pt}
\begin{center}
    \subfloat[{$2000$}]
        {
        	    \label{fig:ex-match-old-2}
                \includegraphics[width=0.25\textwidth]{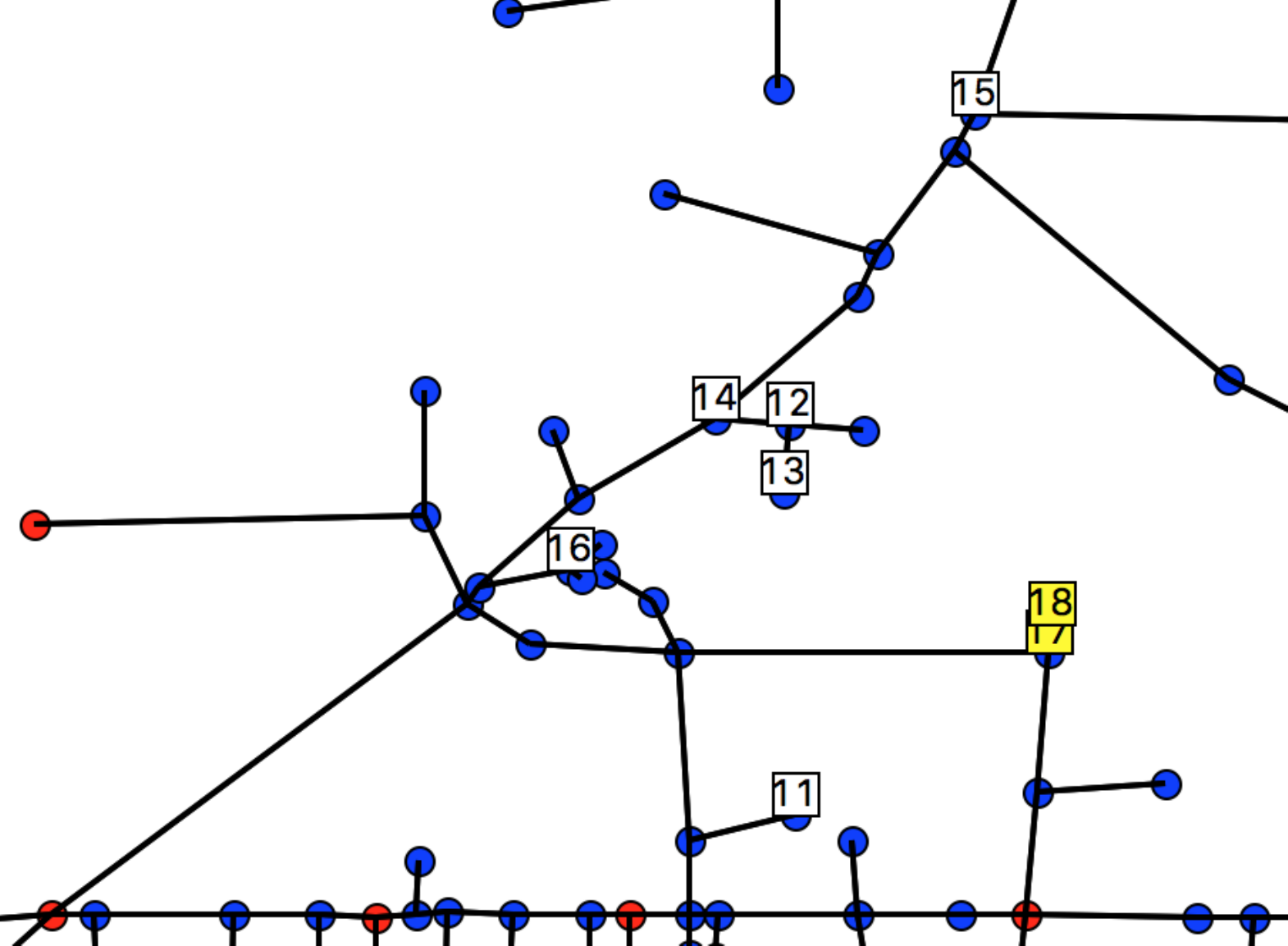}
        }
	\subfloat[{$2006$}]
        {
         	    \label{fig:ex-match-new-2}
                \includegraphics[width=0.25\textwidth]{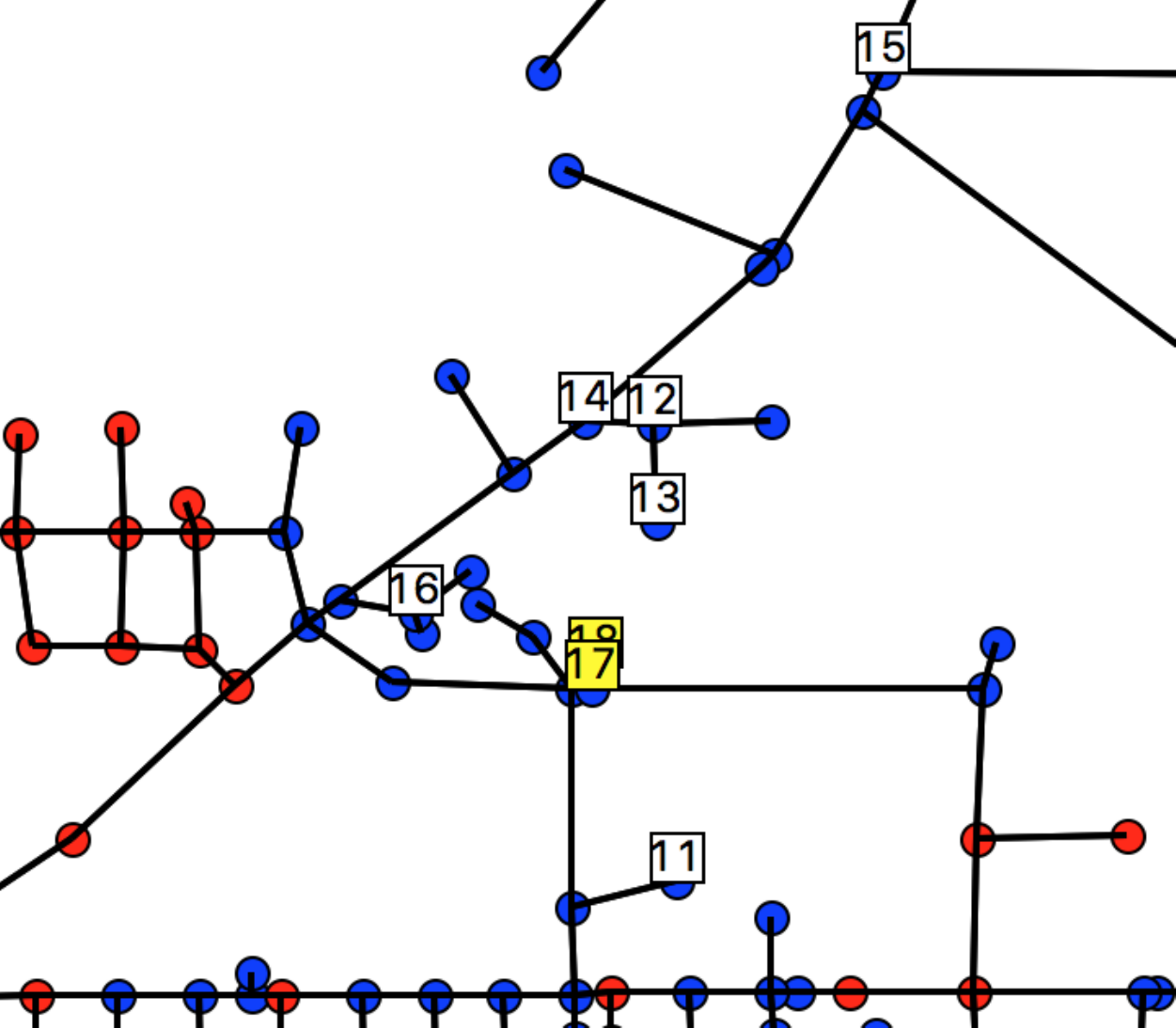}
        }
\end{center}
\vspace*{-12pt}
\caption{Second example of a portion of a matching for Del Norte County, CA where $k = 3$}
\label{fig:ex-matching-2}
\end{figure*}

\begin{figure*}[hbtp]
\vspace*{-12pt}
\begin{center}
    \subfloat[{$2000$}]
        {
                \includegraphics[width=0.2\textwidth]{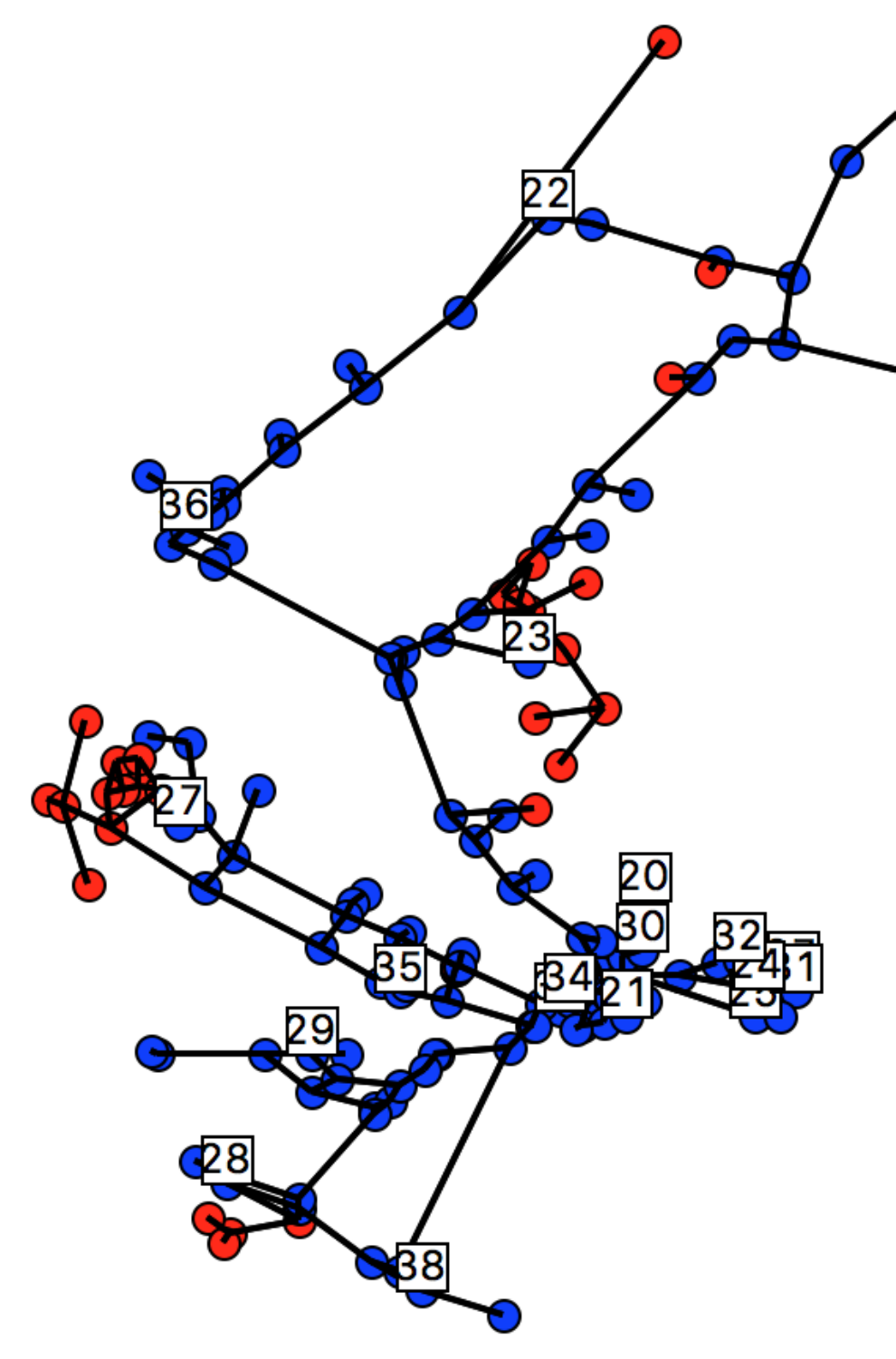}
        }
	\subfloat[{$2006$}]
        {
                \includegraphics[width=0.2\textwidth]{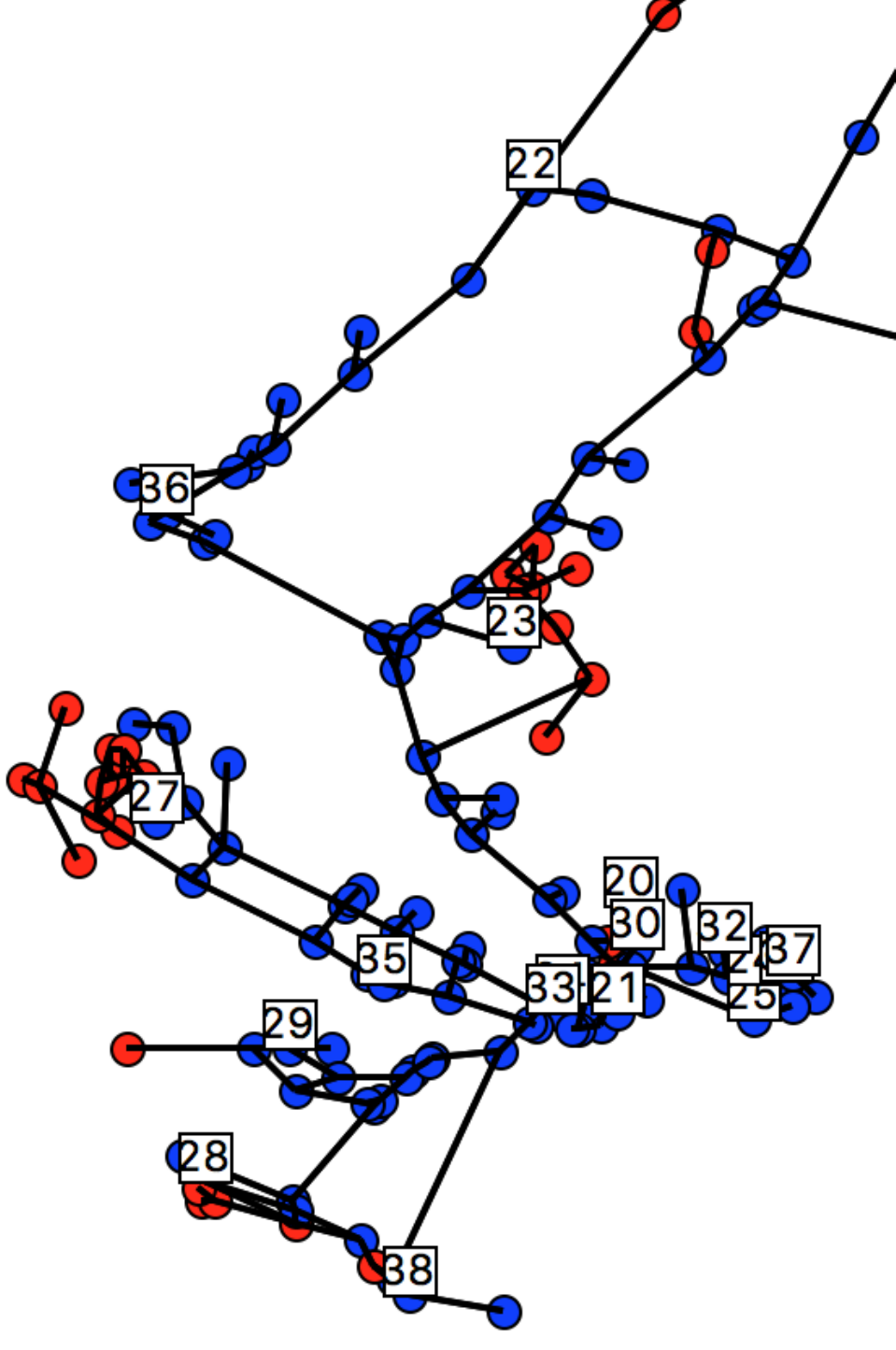}
        }
\end{center}
\vspace*{-12pt}
\caption{Third example of a portion of a matching for Del Norte
County, CA where $k = 3$.}
\label{fig:ex-matching-3}
\end{figure*}

\begin{figure*}[hbtp]
\vspace*{-12pt}
\begin{center}
    \subfloat[{$2000$}]
        {
                \includegraphics[width=0.2\textwidth]{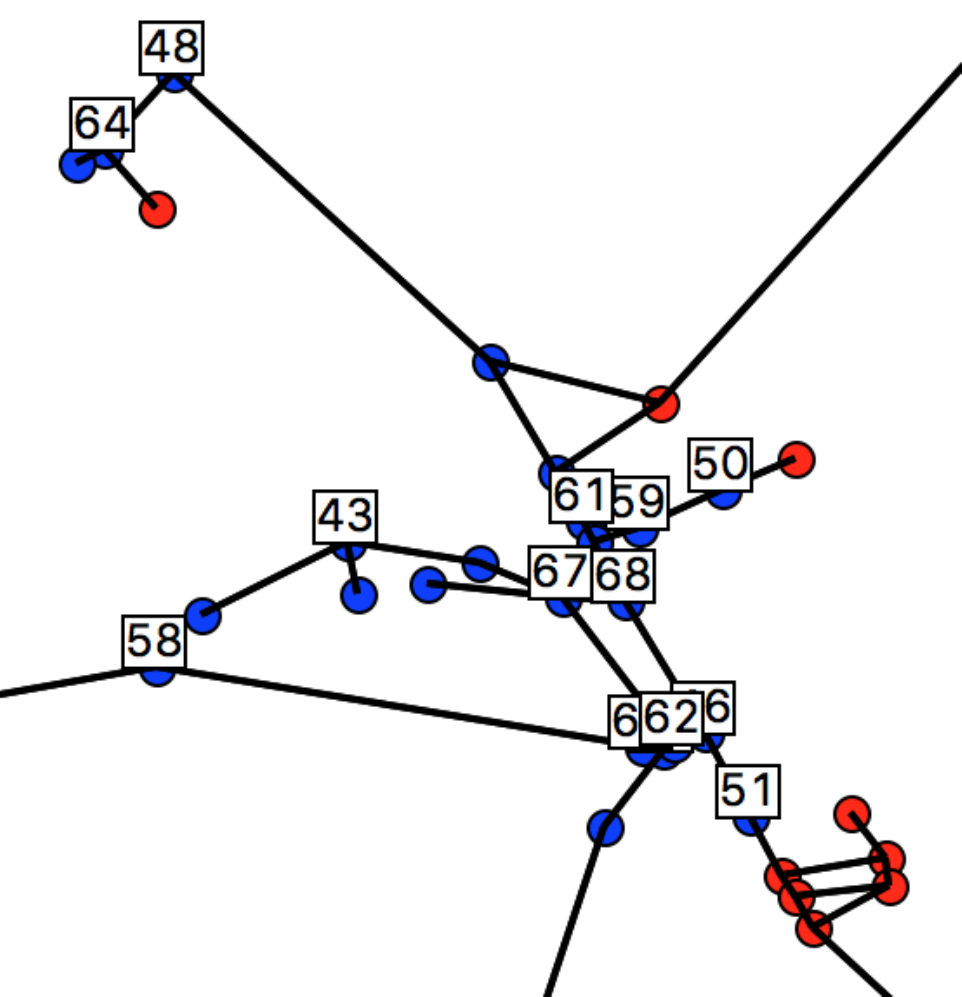}
        }
	\subfloat[{$2006$}]
        {
                \includegraphics[width=0.2\textwidth]{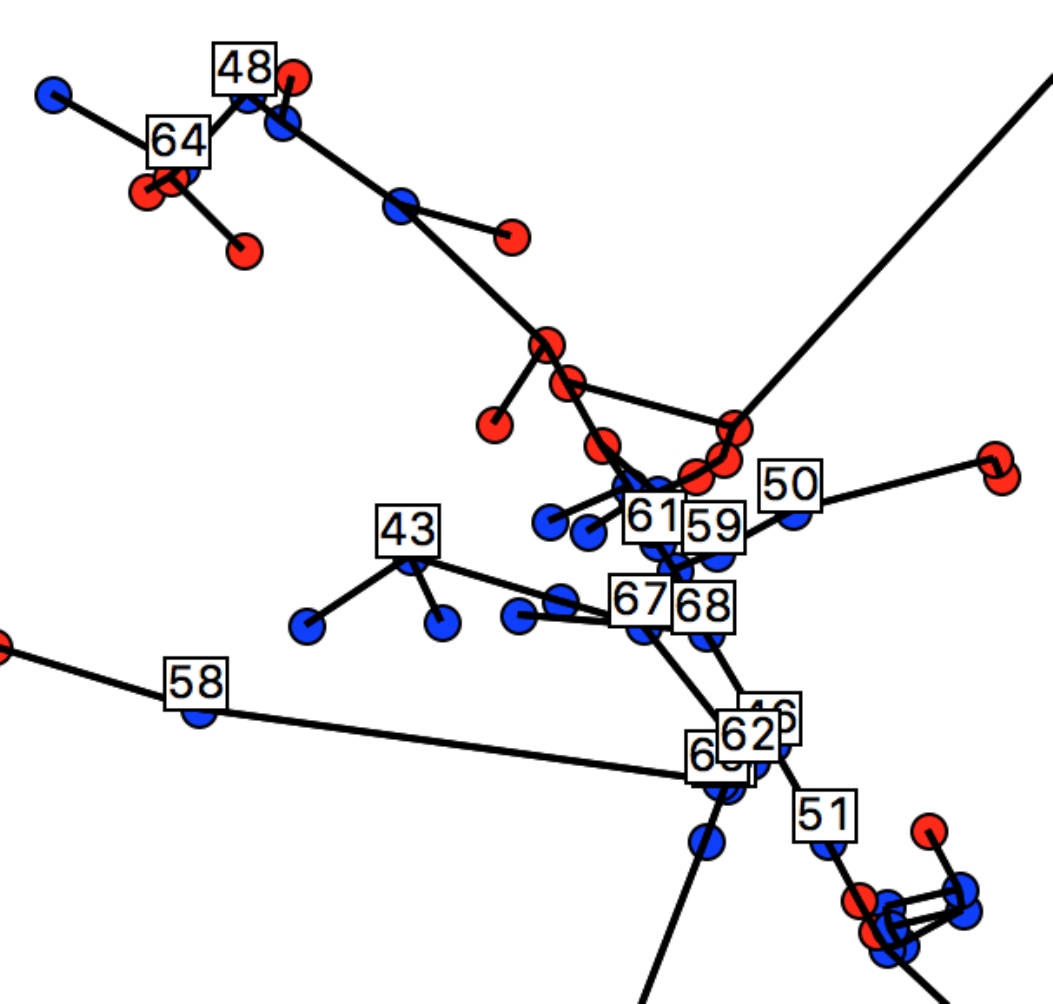}
        }
\end{center}
\vspace*{-12pt}
\caption{Fourth example of a portion of a matching for Del Norte
County, CA where $k = 3$.}
\label{fig:ex-matching-4}
\end{figure*}

In this subsection, we provide a visualization of the matching our
algorithm created for Del Norte County, CA.  We performed the
matching with $k = 3$ and then took four snapshots of the matching
to enlarge the details. For Figures~\ref{fig:ex-matching-1},
\ref{fig:ex-matching-2}, \ref{fig:ex-matching-3}, and
\ref{fig:ex-matching-4}, a node is colored blue if it was matched
and red otherwise.  Furthermore, a node with a white box above it
from the first image containing number $i$ matches the node with a
white box above it containing the number $i$ from the second image.

First, consider Figure~\ref{fig:ex-matching-1}.   Solely based off of geographic location, it is clear that the nodes are being matched to the correct area.  After further inspection, it can be seen that the graph has remained nearly the same around the white boxes containing ``1", ``3", ``7", and ``8".  Near each of these white boxes, our matching algorithm has matched the correct nodes, indicated by all of the blue nodes surrounding said boxes.  Figure~\ref{fig:ex-matching-1} also demonstrates the issue of using a small value for $k$.  The yellow boxes in Figure~\ref{fig:ex-match-new-1} indicate nodes that have been matched to other nodes in the graph from Figure~\ref{fig:ex-match-old-1} that are not included in the image.  This incorrect matching is due to the fact that when $k$ is small, as mentioned earlier, it is likely that many nodes will end up with the same label, yielding a higher likelihood of incorrectly matching two nodes that should not be matched. 

Second, consider Figure~\ref{fig:ex-matching-2}.  The white boxes in these figures are here to indicate that the matching algorithm is performing properly in many parts of the graph.  As we are just using topological features, we also get some unexpected matching as shown in Figure~\ref{fig:ex-matching-2}.  The two yellow boxes in Figure~\ref{fig:ex-match-old-2} are matched to the two yellows boxes in Figure~\ref{fig:ex-match-new-2}.  A new vertex was added in the 2006 graph that caused the matching of the vertices under the yellow boxes to occur in the wrong place.  Because we are only using topological features, our matching algorithm cannot distinguish between the new vertex and the old one that it should be matching to.

Last, consider Figures~\ref{fig:ex-matching-3} and \ref{fig:ex-matching-4}.  Many more white boxes were included to show the success of our matching algorithm in these portions of the graph.


\subsection{Detailed Analysis}

We ran our algorithm on $40$ different counties in California ranging
from small counties to big counties. The results for our
experiments are shown in Table~\ref{tab:overall-results}. Each row
gives analysis about one particular county where $G_1$ is obtained
from TIGER/Line ASCII format from the year $2000$ and $G_2$ is
obtained from TIGER/Line ASCII format from the year $2006$. The
column titled ``seed time" indicates the time taken to find the
seed vertices for the given value of $k$ and the column titled
``match time" indicates the amount of time taken for the topological
flood-based matching algorithm. We ran the experiments on a machine
with $3.1$ GHz Intel Core i7 CPU and $16$ GB of RAM and report the
timings in seconds.  The last column titled ``thresh. ratio" is the
ratio of number of pairs of matched vertices within $5$ miles of
each other to the total number of pairs of matched vertices which
gives up the quality of matching. We can see from the table that
thresh. ratio is always greater than $0.9$ which tells us that our
algorithm performs well on all kinds of inputs.

\begin{table*}[htb!]
\centering
{
\small
\caption{Results for various counties throughout California}
\begin{tabular}{| c | c | c | c | c | c | >{\centering\arraybackslash}p{1.2cm} | >{\centering\arraybackslash}p{1.45cm} | c | c |} \hline
county & $k$ & nodes of $G_1$ & nodes of $G_2$ & edges of $G_1$ & edges of $G_2$ & seed time (seconds) & match time (seconds) & approx. ratio & thres. ratio \\ \hline
Alameda & 9 & 40752 & 40242 & 67226 & 66644 & 778.544 & 7.686 & 0.9777 & 0.9997\\ \hline 
Alpine & 5 & 1448 & 1427 & 1838 & 1811 & 0.536 & 0.311 & 0.9439 & 0.9985 \\ \hline 
Amador & 5 & 6970 & 6784 & 9198 & 8991 & 3.017 & 0.764 & 0.9553 & 0.9998\\ \hline 
Butte & 8 & 19856 & 21304 & 27896 & 29955 & 66.540 & 34.509 & 0.6878 & 0.9963\\ \hline 
Calaveras & 8 & 13770 & 13043 & 18141 & 17690 & 23.050 & 1.185 & 0.0400 & 0.9196\\ \hline 
Colusa & 5 & 5039 & 5700 & 7285 & 8589 & 4.231 & 8.106 & 0.8106 & 0.9867\\ \hline 
Contra Costa & 9 & 37148 & 36564 & 55555 & 54750 & 323.512 & 70.794 & 0.9482 & 0.9995\\ \hline 
Del Norte & 5 & 5383 & 7034 & 7386 & 9785 & 3.861 & 39.562 & 0.4811 & 0.9258\\ \hline 
El Dorado & 9 & 24248 & 24103 & 33331 & 33271 & 108.336 & 10.559 & 0.9766 & 0.9991\\ \hline 
Fresno & 9 & 51006 & 50614 & 83081 & 82640 & 744.009 & 346.011 & 0.9796 & 0.9992\\ \hline 
Imperial & 8 & 18104 & 18105 & 28716 & 28639 & 92.026 & 1.481 & 0.9592 & 1.0\\ \hline 
Kings & 8 & 11842 & 15328 & 18775 & 25521 & 87.083 & 7.071 & 0.3541 & 0.9978\\ \hline 
Lake & 8 & 12437 & 18500 & 17486 & 26176 & 47.485 & 391.934 & 0.1331 & 0.9553\\ \hline 
Lassen & 8 & 16216 & 19519 & 24024 & 28044 & 57.645 & 241.791 & 0.3804 & 0.9837\\ \hline 
Madera & 8 & 16936 & 16633 & 25164 & 24842 & 77.837 & 4.864 & 0.9633 & 0.9998\\ \hline 
Marin & 8 & 13733 & 13455 & 19722 & 19372 & 44.946 & 1.231 & 0.9446 & 1.0\\ \hline 
Mariposa & 5 & 9241 & 10538 & 12033 & 13830 & 4.558 & 119.975 & 0.5746 & 0.9546\\ \hline 
Mendocino & 9 & 22326 & 26153 & 30231 & 36142 & 107.554 & 403.188 & 0.3761 & 0.9901\\ \hline 
Merced & 8 & 16576 & 19619 & 25266 & 29568 & 78.726 & 9.331 & 0.6058 & 0.9940\\ \hline 
Modoc & 8 & 13304 & 17408 & 19674 & 24889 & 44.782 & 3.998 & 0.3229 & 0.9837\\ \hline 
Mono & 5 & 9159 & 11345 & 13203 & 16178 & 6.440 & 30.914 & 0.6179 & 0.9595\\ \hline 
Monterey & 9 & 31887 & 33831 & 48204 & 51313 & 324.838 & 350.278 & 0.7654 & 0.9978\\ \hline 
Napa & 5 & 6932 & 6827 & 10054 & 9867 & 4.813 & 19.717 & 0.9491 & 0.9933\\ \hline 
Nevada & 8 & 15903 & 15268 & 21729 & 20935 & 30.650 & 6.726 & 0.9135 & 0.9994\\ \hline 
Placer & 9 & 25365 & 25437 & 35603 & 35913 & 116.914 & 41.591 & 0.9287 & 0.9996\\ \hline 
San Benito & 5 & 7555 & 10311 & 10421 & 14649 & 5.801 & 30.310 & 0.5597 & 0.9503\\ \hline 
San Francisco & 5 & 9803 & 11570 & 20313 & 24218 & 20.241 & 0.977 & 0.7138 & 1.0\\ \hline 
San Mateo & 9 & 21571 & 21101 & 35132 & 34532 & 288.204 & 5.469 & 0.9666 & 0.9997\\ \hline 
Santa Cruz & 8 & 14374 & 14063 & 20545 & 20142 & 48.941 & 19.229 & 0.9694 & 0.9997\\ \hline 
Shasta & 9 & 25436 & 33824 & 35129 & 47588 & 163.593 & 431.638 & 0.1867 & 0.9732\\ \hline 
Sierra & 5 & 4809 & 6522 & 6603 & 8912 & 3.273 & 5.000 & 0.3559 & 0.8603\\ \hline 
Siskiyou & 9 & 28210 & 38150 & 39682 & 52616 & 140.771 & 705.416 & 0.1442 & 0.9528\\ \hline 
Solano & 8 & 15930 & 31249 & 24859 & 46954 & 106.300 & 63.848 & 0.1398 & 0.9654\\ \hline 
Stanislaus & 8 & 18254 & 19240 & 29327 & 31094 & 110.927 & 1.703 & 0.5786 & 0.9962\\ \hline 
Sutter & 5 & 6311 & 6164 & 9670 & 9494 & 4.293 & 0.319 & 0.9704 & 1.0\\ \hline 
Tehama & 8 & 15399 & 19177 & 21756 & 27353 & 53.793 & 13.337 & 0.3802 & 0.9844\\ \hline 
Trinity & 8 & 12042 & 11944 & 15501 & 15434 & 15.961 & 77.516 & 0.9780 & 0.9945\\ \hline 
Tulare & 9 & 27555 & 27302 & 42308 & 42257 & 269.419 & 10.146 & 0.9632 & 0.9993\\ \hline 
Tuolumne & 9 & 14830 & 17135 & 19985 & 23283 & 47.406 & 131.849 & 0.2621 & 0.9809\\ \hline 
Yuba & 5 & 9354 & 9267 & 13529 & 13407 & 5.791 & 4.158 & 0.9840 & 0.9995\\ \hline

\end{tabular}
\label{tab:overall-results}
}
\end{table*}

\begin{figure}
\begin{center}
    {
        \includegraphics[width=\columnwidth]{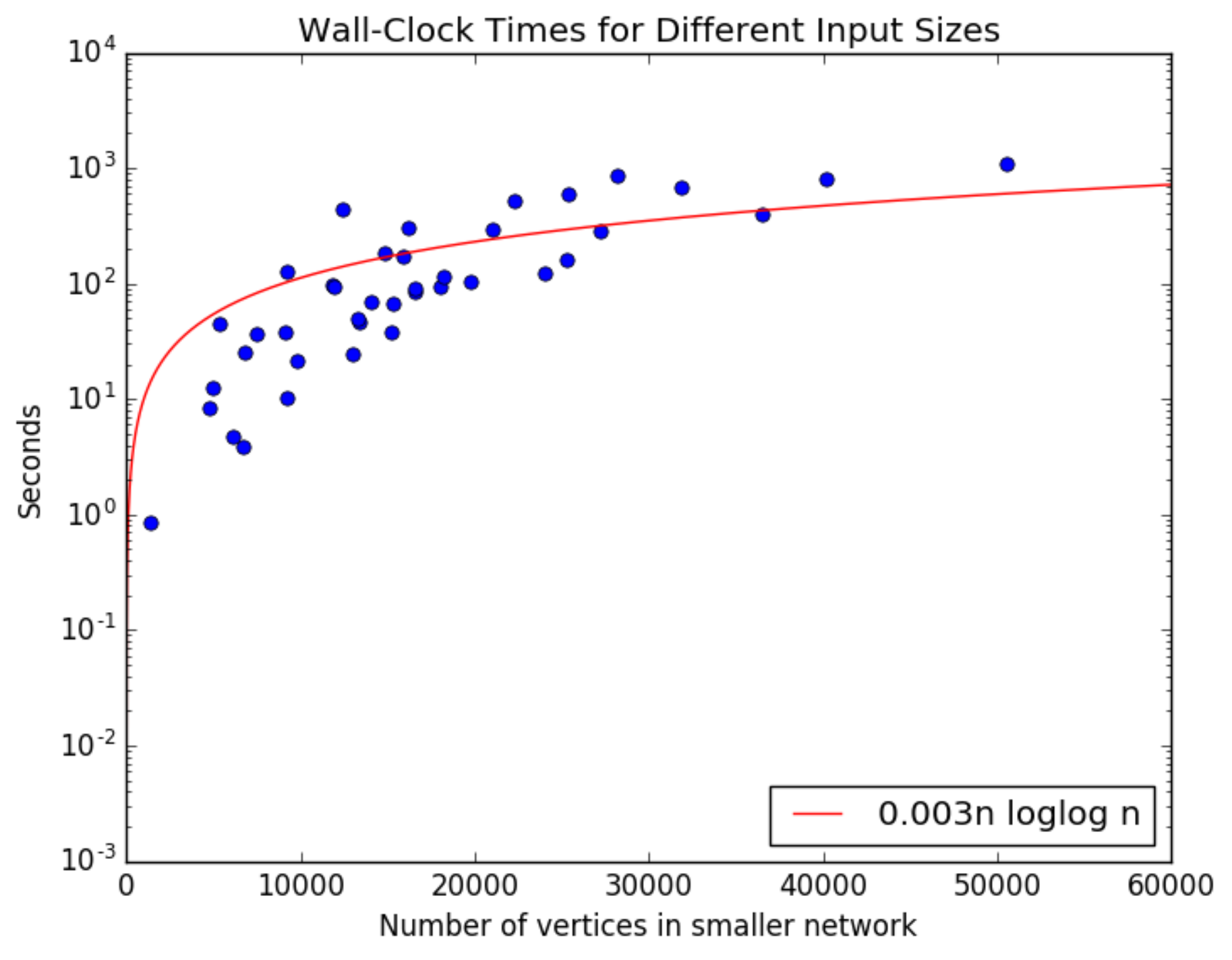}
    }
\end{center}
\caption{Running times for our algorithm on the graphs given in Table~\ref{tab:overall-results}} 
\label{fig:runtime-against-size}
\end{figure}

Figure~\ref{fig:runtime-against-size} plots the experiments in Table~\ref{tab:overall-results} with the total running time (seed time plus match time) as the $y$-axis and the size of the smaller graph as the $x$-axis.  The red line represents the function $0.003n\log\log n$. Therefore, it seems that the variable $C$ defined in Section 2 is much less than $n$, which is good for the running time of our algorithm.

\section{Conclusion}
We have given a purely topological algorithm for determining the
changes that occur between two road networks, and we have provided
both theoretical and experimental analysis to show that our algorithm
is effective and efficient.
We therefore feel that this algorithm provides a good tool for
solving the map evolution problem when geometric or geographic
features are missing from one or both of the road networks being
considered.

\subsection*{Acknowledgments}
This article reports on work supported by the Defense Advanced
Research Projects Agency under agreement no.~AFRL FA8750-15-2-0092.
The views expressed are those of the authors and do not reflect the
official policy or position of the Department of Defense or the
U.S.~Government.
This work was also supported in part by the U.S.~National Science Foundation
under grants 1228639 and 1526631.
We would like to thank David Eppstein for several helpful
discussions related to the topics of this paper.

\bibliographystyle{abbrv}
\bibliography{matching}

\end{document}